%% file: main.tex
\documentclass[lettersize,journal]{IEEEtran}
\usepackage{amsmath,amsfonts}
\usepackage{array}
\usepackage[caption=false,font=normalsize,labelfont=sf,textfont=sf]{subfig}
\usepackage{textcomp}
\usepackage{stfloats}
\usepackage{url}
\usepackage{verbatim}
\usepackage{graphicx}
\def\BibTeX{{\rm B\kern-.05em{\sc i\kern-.025em b}\kern-.08em
T\kern-.1667em\lower.7ex\hbox{E}\kern-.125emX}}
\usepackage{balance}
\input{./meta/commands}

\begin{document}

\title{LightVA: \revised{Lightweight} Visual Analytics with \revised{LLM} Agent-Based Task Planning \revised{and Execution} }

\input{./meta/authors}

\markboth{Journal of \LaTeX\ Class Files,~Vol.~18, No.~9, September~2020}%
{How to Use the IEEEtran \LaTeX \ Templates}

\maketitle

\input{sections/0-abstract}

\begin{IEEEkeywords}
Visual Analytics, Task Planning, Large Language Model Agent, Mixed-Initiative Interaction
\end{IEEEkeywords}

\begin{spacing}{0.88}
\input{sections/1-introduction}
\input{sections/2-related-work}
\input{sections/3-requirement-analysis}
\input{sections/3.1-workflow-framework}
\input{sections/4-flow-model}

\input{sections/5-framework}
\input{sections/6-usage-scenario}

\input{sections/7-user-study}

\input{sections/7.1-results}

\input{sections/8-discussion}

\input{sections/9-conclusion}
\end{spacing}

\section*{Acknowledgments}
We thank anonymous reviewers for their constructive comments. This work is supported by the Natural Science Foundation of China (NSFC No.62472099).

\bibliographystyle{IEEEtran}
\bibliography{main}

\newpage
\appendices
\input{sections/appendix}

\input{sections/appendix-test-llm}

\end{document}

%% file: meta/commands.tex
\newcommand{\revised}[1]{{\color{black} #1}}
\newcommand{\rerevised}[1]{{\color{black} #1}}

\newcommand{\human}[1]{{\color{human}\textbf{#1}}}
\newcommand{\agent}[1]{{\color{agent}\textbf{#1}}}

\usepackage{cite}                      

\usepackage{hyperref}
\usepackage{cleveref}
\Crefname{figure}{Fig.}{Figs.}  
\usepackage{tabularx}
\usepackage[svgnames]{xcolor} 
\definecolor{human}{HTML}{958857}
\definecolor{agent}{HTML}{BA5252}
\definecolor{comm}{gray}{0.6}
\definecolor{comm}{gray}{0.6}
\definecolor{NavyBlue}{RGB}{0,0,128} 

\usepackage{algorithm}
\usepackage{algpseudocodex}

\usepackage{amssymb,amsmath,amsthm,enumitem}
\usepackage{mathptmx} 

\algrenewcommand\algorithmicrequire{\textbf{Input:}}
\algrenewcommand\algorithmicensure{\textbf{Output:}}
\algnewcommand{\IfThen}[2]{\State \algorithmicif\ #1\ \algorithmicthen\ #2}
\newcommand{\method}[0]{\textsc{TPE}}
\newcommand{\fullform}[0]{\textbf{T}ask \textbf{P}lanning and \textbf{E}xecution}

\usepackage{tabu}                      
\usepackage{booktabs}                  
\usepackage{mwe}                       

\usepackage{thmtools}
\usepackage[framemethod=TikZ]{mdframed}
\mdfsetup{skipabove=1em,skipbelow=0em}

\newcommand{\schemaname}[1]{{\small\fontfamily{txtt}\selectfont #1}}

\PassOptionsToPackage{svgnames}{xcolor}
\usepackage{xcolor}
\usepackage{xspace, xpunctuate}
\usepackage{listings}  
\usepackage{enumitem}  
\usepackage{multirow}
\definecolor{keywordcolor}{rgb}{0.56, 0.13, 0.00}
\definecolor{ndkeywordcolor}{rgb}{0.05, 0.46, 0.17}
\definecolor{commentcolor}{rgb}{0.41, 0.64, 0.70}
\definecolor{stringcolor}{rgb}{0.25, 0.44, 0.63}
\lstdefinelanguage{TypeScript}{
  keywords={typeof, new, true, false, catch, function, return, null, catch, switch, var, if, in, while, do, else, case, break, boolean},
  morekeywords={[2]{class, export, throw, implements, import, this}},
  identifierstyle=\color{black},
  sensitive=false,
  comment=[l]{//},
  morecomment=[s]{/*}{*/},
  commentstyle=\color{commentcolor}\ttfamily,
  stringstyle=\color{stringcolor}\ttfamily,
  morestring=[b]',
  morestring=[b]"
}
\lstdefinelanguage{Python}{
  keywords={def, class, if, elif, else, while, for, break, continue, return, from, import, as, pass, raise, with, in, not, and, or, is, None, True, False, try, except, finally},
  morekeywords={[2]{@staticmethod, @classmethod, @property}},
  identifierstyle=\color{black},
  sensitive=true,
  comment=[l]{\#},
  morecomment=[s]{'''}{'''},
  morecomment=[s]{"""}{"""},
  commentstyle=\color{commentcolor}\ttfamily,
  stringstyle=\color{stringcolor}\ttfamily,
  morestring=[b]',
  morestring=[b]",
  morestring=[b]"""
}
\lstdefinelanguage{mylang}{
}
\lstdefinestyle{mystyle}{
    basicstyle=\fontfamily{txtt}\selectfont\footnotesize,
    numbers=none,
  frame=none,
  columns=flexible,
  xleftmargin=0pt,
  aboveskip=0pt,
  belowskip=0pt,
  language=mylang
}

\colorlet{punct}{red!60!black}
\definecolor{delim}{RGB}{20,105,176}
\colorlet{numb}{magenta!60!black}

\lstdefinelanguage{json}{
    basicstyle=\fontfamily{txtt}\selectfont\footnotesize,
    numbers=left,
    numberstyle=\color{gray}\footnotesize\ttfamily,
    numbersep=4pt,
    tabsize=2,
    showstringspaces=false,
    breaklines=true,
    xleftmargin=10pt,
    literate=
     *{0}{{{\color{numb}0}}}{1}
      {1}{{{\color{numb}1}}}{1}
      {2}{{{\color{numb}2}}}{1}
      {3}{{{\color{numb}3}}}{1}
      {4}{{{\color{numb}4}}}{1}
      {5}{{{\color{numb}5}}}{1}
      {6}{{{\color{numb}6}}}{1}
      {7}{{{\color{numb}7}}}{1}
      {8}{{{\color{numb}8}}}{1}
      {9}{{{\color{numb}9}}}{1}
      {:}{{{\color{punct}{:}}}}{1}
      {,}{{{\color{punct}{,}}}}{1}
      {\{}{{{\color{delim}{\{}}}}{1}
      {\}}{{{\color{delim}{\}}}}}{1}
      {[}{{{\color{delim}{[}}}}{1}
      {]}{{{\color{delim}{]}}}}{1},
}

\definecolor{mycolor1}{RGB}{147,137,93}

\definecolor{mycolor}{RGB}{147,137,93}
\definecolor{bgcolor}{RGB}{251,249,246}

\usepackage{setspace}
\declaretheoremstyle[
  headfont=\small\bfseries\sffamily\color{mycolor},
  postheadspace=\newline,
  mdframed={
      linewidth=2pt,
      leftline=false, rightline=false, topline=false, bottomline=false,
      linecolor=bgcolor, backgroundcolor=bgcolor,      innerleftmargin=5pt, 
      innerrightmargin=5pt, 
  }
]{thmgreenbox}

\declaretheoremstyle[
    headfont=\bfseries\sffamily\color{NavyBlue!70!black}, bodyfont=\normalfont,
    mdframed={
        linewidth=2pt,
        rightline=false, topline=false, bottomline=false,
        linecolor=NavyBlue, backgroundcolor=NavyBlue!5,
    }
]{thmbluebox}

\declaretheoremstyle[
    headfont=\bfseries\sffamily\color{NavyBlue!70!black}, bodyfont=\normalfont,
    numbered=no,
    mdframed={
        linewidth=2pt,
        rightline=false, topline=false, bottomline=false,
        linecolor=NavyBlue, backgroundcolor=NavyBlue!1,
    },
]{thmexplanationbox}

\declaretheorem[style=thmgreenbox, name=Prompt Template]{Prompt}

\declaretheorem[style=thmexplanationbox, name=Proof]{tmpexplanation}

%% file: meta/authors.tex
\author{Yuheng~Zhao, Junjie Wang, Linbin Xiang, Xiaowen Zhang, Zifei Guo,\\ Cagatay~Turkay, Yu Zhang and Siming~Chen

\thanks{Yuheng~Zhao, Junjie Wang, Linbin Xiang, Xiaowen Zhang, Zifei Guo, Siming Chen are with School of Data Science, Fudan University. E-mail: \{yuhengzhao, simingchen\}@fudan.edu.cn. Siming Chen is the corresponding author.} 
\thanks{Cagatay Turkay is with the Centre for Interdisciplinary Methodologies, University of Warwick. 
E-mail: Cagatay.Turkay@warwick.ac.uk.}
\thanks{Yu Zhang is with Department of Computer Science, University of Oxford. E-mail: yuzhang94@outlook.com.
}
\thanks{Manuscript received April 19, 2005; revised August 26, 2015.}}


%% file: sections/0-abstract.tex
\begin{abstract}
Visual analytics (VA) requires analysts to iteratively propose analysis tasks based on observations and execute tasks by creating visualizations and interactive exploration to gain insights.
This process demands skills in programming, data processing, and visualization tools, highlighting the need for a more intelligent, streamlined VA approach.
\rerevised{Large language models (LLMs) have recently been developed as agents to handle various tasks with dynamic planning and tool-using capabilities, offering the potential to enhance the efficiency and versatility of VA.}
We propose LightVA, a lightweight VA framework that supports task decomposition, data analysis, and interactive exploration through human-agent collaboration.
Our method is designed to help users progressively translate high-level analytical goals into low-level tasks, producing visualizations and deriving insights.
Specifically, we introduce an LLM agent-based task planning and execution strategy, employing a recursive process involving a planner, executor, and controller. 
The planner is responsible for recommending and decomposing tasks, the executor handles task execution, including data analysis, visualization generation and multi-view composition, and the controller coordinates the interaction between the planner and executor.
\rerevised{Building on the framework, we develop a system with a hybrid user interface that includes a task flow diagram for monitoring and managing the task planning process, a visualization panel for interactive data exploration, and a chat view for guiding the model through natural language instructions.}
We examine the effectiveness of our method through a usage scenario and an expert study.
\end{abstract}

%% file: sections/1-introduction.tex
\section{Introduction}
\label{sec:introduction}

Visual analytics (VA) deciphers complex datasets with data mining and interactive visualizations~\cite{Thomas2005Illuminating, Keim2008Visual}. 
\rerevised{However, building} and using a VA system \rerevised{can be} a costly endeavor that encompasses several main stages: goal understanding, task decomposition, data modeling, and visualization creation to discover insights.
A key challenge is that this process is iterative, requiring continual refinement based on evolving needs~\cite{Wu2023Defence}. 
Different tasks necessitate various data analysis and visualization methods to form a VA system. While using the system, tasks may evolve based on the insights gained, necessitating ongoing iterations until the analytical goals are achieved~\cite{ceneda2016characterizing}.
Consider a scenario where the goal is to identify high-risk events from social media data.
\rerevised{Users may first need an overview from different perspectives, such as the distribution of keywords over time or changes in sentiment for risk analysis. If outliers are identified, users may need further details, such as spatial distribution or entity relationships. In this process, the task space is broad and fluid, requiring efficient task planning and method implementation.}

Recent research focuses on data-driven or natural language-based visual data exploration, with an emphasis on automatic visualization generation~\cite{Deng2023DashBot, yu2019flowsense} or insight mining~\cite{Wang2020DataShot,demiralp2017foresight}. 
Large Language Models (LLMs) present a potential for data analysis, supporting dynamic task planning and lower development costs across various scenarios. 
The reasoning abilities that enable autonomous planning and execution of analytical tasks~\cite{hao2023reasoning, sharan2023llm,yang2024intercode}, while code generation capability support creating insightful visualizations efficiently~\cite{liu2023jarvix, dibia2023lida, wang2023llm4vis}.
Additionally, their broad knowledge base makes LLMs versatile tools capable of adapting to diverse data analysis contexts~\cite{mao2023gpteval, zhang2023data}.
LEVA~\cite{zhao2024LEVA} integrates LLMs in VA systems to recommend insights for a given task but still cannot generate visualization and data modeling methods adapted to tasks. 
\revised{There is a lack of approaches supporting task planning, VA methods implementation, and interactive analysis with human-agent collaboration.}

This paper introduces \textit{LightVA}, a lightweight VA framework with agent-based task planning.
\revised{The term ``lightweight'' refers to the framework's focus on reducing the cost of development and using VA systems.
Using LLM agents to aid the task planning and execution process.}
\rerevised{The framework builds upon multi-level relationships, translating high-level goals to low-level tasks and deriving insights through data mining and interactive visualizations.}
Specifically, the framework employs a recursive process that includes a planner, executor, and controller, which dynamically accommodates task complexity.
The planner is responsible for task decomposition, the executor handles task execution, including visualization generation and data analysis, and the controller orchestrates the executor and planner and manages whether tasks continue to be decomposed.
\revised{We develop a system based on the framework that provides a chat view to support communication among users and agents, a task flow view to visualize manage the process of task planning, a visualization panel to show single view and multiple linked views connected to the task flow.}
Our main contributions are as follows: 
\begin{itemize}[leftmargin=*]
    \item We propose a lightweight VA framework using LLM agent-based task planning and execution. This approach enables adaptive, efficient analysis through human-agent collaboration, supporting users in task decomposition, visualization, and insight discovery.
    \item We develop a system that embodies our framework, supporting users to analyze data with the assistance of agents and communicate through the hybrid user interface. 
    \item We demonstrate the effectiveness of the system through a usage scenario and an expert study. 
\end{itemize}

%% file: sections/2-related-work.tex
\section{Related Work}
\label{sec:related-work}

Our research is related to prior studies on visualization recommendations, task-driven data exploration, and LLM application in data exploration.

\subsection{Visualization Recommendation}

Visualization authoring typically requires users to have professional visualization knowledge and programming ability. 
For example, tools like Tableau support creating visualizations and multiple linked views with shelf-configuration design, offering robust visualization creation features.
However, while Tableau excels as an authoring tool, it offers limited support for automatic task decomposition and VA method implementation to further assist with analysis.
A corpus of research has been proposed on automatic visualization recommendations~\cite{wu2021ai4vis, wang2021survey, ren2023re}. 
Rule-based methods, such as Voyager~\cite{wongsuphasawat2015voyager} and CompassQL~\cite{wongsuphasawat2016towards}, utilize the visualization principles to construct visual mapping and allow users to choose their interested data properties and visual encoding to create visualizations.
\revised{For machine learning methods, Data2Vis~\cite{dibia2019data2vis} introduces an end-to-end trainable neural translation model for automatically generating visualizations from given datasets. 
VizML~\cite{Hu2019VizML} learned visualization design choices from a corpus of data-visualization pairs. 
Table2Charts~\cite{zhou2021table2charts} recommends visualizations by learning patterns between tables and visualizations.
ChartSeer~\cite{zhao2020chartseer} employs deep learning to recommend visualizations based on users' interactions.
Unlike end-to-end deep learning methods that directly learn from datasets to generate visualizations, knowledge graph-based approaches, such as AdaVis~\cite{zhang2023adavis}, KG4VIS~\cite{li2021kg4vis}, Lodestar~\cite{raghunandan2022lodestar}, leverage structured information about data and relationships to recommend visualizations.}

In addition, there are some works that consider multi-view generation.
\rerevised{Qu and Hullman~\cite{qu2017keeping} proposes coordination principles to keep consistency. Sun et al.\cite{sun2021towards} investigate different linking techniques based on data relationships.}
Dziban~\cite{lin2020dziban} is a visualization API using anchored recommendation and extending Draco~\cite{moritz2018formalizing} to reason about multiple views. 
DMiner~\cite{Lin2023DMiner} investigated the design rules of the single views and view-wise relationships from online notebooks to recommend multiple-view dashboards.
MultiVision~\cite{wu2021multivision} and DashBot~\cite{Deng2023DashBot} recommend dashboards given an input dataset in an end-to-end manner using deep learning models.
\revised{Shi et al.~\cite{shi2023reverse} optimize multi-view layouts by predicting the similarity of visual elements using Transformer-based models.}
\rerevised{Previous work has provided a solid research foundation for the principles between data, visualization, and multi-views. Building on this, we further study integrating LLM-agent to recommend visualizations that align with high-level goals and evolving tasks throughout the VA pipeline, which require significant human effort.}

\subsection{Task-Driven Visual Data Exploration}

In addition to data attributes when recommending visualizations, some visualization recommendation systems are task-driven recommendation systems that consider one or more analytic tasks (e.g., correlate, analyze trend). 
\rerevised{Some scholars study the recommendation of analysis methods in exploratory data analysis (EDA) within notebooks. EDAssistant~\cite{li2023edassistant} recommends code by analyzing associations between APIs in a large notebook collection. 
ATENA~\cite{bar2020automatically} shapes EDA into a Markov Decision Process (MDP) model using a deep reinforcement learning architecture to effectively optimize notebook generation. 
Furthermore, visual analytics incorporates visualization techniques into the EDA process, expanding the task space.}
Casner~\cite{casner1991task} presents one of the earliest examples of visualization systems that suggest charts based on a user’s task (e.g., finding direct flight routes or a table to see flight information).
Saket et al.~\cite{saket2018task} conducted a study to assess the effectiveness of five canonical visualizations on ten low-level analytic tasks~\cite{amar2005low} and developed a recommendation engine based on their study’s findings. 
Gotz and Wen~\cite{gotz2009behavior} present a prototype system that observes interaction patterns (e.g., repeatedly changing filters or swapping attributes) to infer analytic tasks such as comparison or trend analysis and correspondingly recommends visualizations such as small multiples or line charts. 
VizAssist~\cite{bouali2016vizassist} enables its users to specify their data objectives in terms of analytic tasks (e.g., correlate, compare) and considers these tasks in combination with existing perceptual guidelines as input to a genetic algorithm for recommending visualizations. 
Foresight~\cite{demiralp2017foresight} uses tasks like distributions, outliers, and correlations to guide insight discovery and grouping recommendations.
TaskVis~\cite{shen2021taskvis} recommends visualizations under specific tasks through answer set programming.

In addition to visual generation based on a single task, Medley~\cite{pandey2022medley} recommends multi-view collections based on several analytic intents, and views and widgets can be selected to compose a variety of dashboards.
\rerevised{However, the analytic intents and visualization combinations are often chosen from preset options, which limits exploration flexibility.}
Different from them, we study dynamic task planning based on the goal and findings and leverage human-agent collaboration.

\subsection{Large Language Model in Data Exploration}

LLM-based tools have been proposed for data exploration and visualization tasks.
For visualization generation and recommendations,
LLM4Vis~\cite{wang2023llm4vis} and ChartGPT~\cite{ChartGPT} utilize LLMs to choose appropriate visualizations from natural language instructions. 
Li et al.~\cite{li2024visualization} evaluate the capability of GPT-3.5 to generate visualization specifications, demonstrating its superiority over previous machine learning-based approaches.
NL2Rigel~\cite{huang2023interactive} showcases the LLM's ability to convert instructions into comprehensive data visualizations and tables.
For analytical task translation and automation, Hassan et al.\cite{hassan2023chatgpt} and Data-Copilot\cite{zhang2023data} concentrate on converting analytical goals and ambiguous queries into actionable data analysis tasks.
Ma et al.\cite{ma2023demonstration} and JarviX\cite{liu2023jarvix} introduce systems that automate the data exploration process by identifying suitable analysis intents and generating insights.
Text2Analysis~\cite{he2023text2analysis} offers a framework for categorizing data analysis tasks, establishing a structured approach to tackling common analytical challenges ranging from basic operations to forecasting and chart generation.
However, the tasks in these studies are generally straightforward and focused, with limited exploration into the decomposition of more complex tasks.

When using LLMs to solve complex tasks where multi-step reasoning is demanded, the performance of directly using LLMs tends to decrease.
Recently, prior methods have utilized LLMs with input-output prompting, CoT~\cite{wei2022chain}, ToT~\cite{yao2024tree} or GoT~\cite{besta2023graph} to perform complex task planning and execution.
These methods proved that LLMs are good at task planning but require appropriate prompting techniques.
Another way is to integrate LLMs in the interface, allowing chaining multiple prompts to address a much wider range of human tasks.
Wu et al.~\cite{wu2022ai} introduce the chaining of AI models, where a complex task is divided into multiple steps.
Talk2Data~\cite{guo2021talk2data} presents a natural language interface that enables users to explore visual data through question decomposition.
However, the linked visualization and more advanced data analysis methods remain limited. 
To enhance this capability, we leverage LLMs and propose an agent-based autonomous task-planning strategy for adaptive VA system construction and exploration.

%% file: sections/3-requirement-analysis.tex
\section{LightVA Framework}
\label{sec:framework}

\revised{

The pipeline of VA involves two stages: development and analysis.
Thus, in LightVA, we aim to reduce the efforts for both developers and analysts.}
In the following, we will examine the challenges and derive the design requirements for integrating LLM-based agents into the user's workflow.
Finally, we introduce the conceptual framework of agent-based VA workflow.

\subsection{Challenges}

Through a review of visual analytics literature, we identify and dissect specific challenges that intensify the effort users must exert:

\begin{enumerate}[leftmargin=6mm]

\revised{\item[\textbf{C1}] \textbf{Accommodating diverse analysis requirements:}}
Data Analysts often face the immense challenge of navigating a vast exploration space, where the analytical process is dynamic and iterative~\cite{ceneda2016characterizing}. Forming hypotheses and validating them through the continuous proposal of new tasks and insights is complex. They have to figure out the connections between tasks and insights in their mind and try to propose tasks in the next few steps until they achieve the goal.

\revised{\item[\textbf{C2}] \textbf{Developing visualization and data mining tools is time-consuming:}}
\revised{Analysts often lack proficiency in developing and managing VA systems, which makes it challenging for them to select and apply appropriate data modeling and visualization techniques~\cite{pandey2022medley}.} Additionally, when tasks involve multiple visualizations, these must be integrated into a linked view to enable more effective interactive exploration~\cite{wang2000guidelines}. However, this process requires substantial knowledge of both visualization principles and coding skills, resulting in inefficiencies and delays~\cite{khan2022rapid}.

\end{enumerate}

\subsection{Design Requirements}
\label{sec:requirements}

\revised{Based on these challenges}, we have settled on a set of targeted design requirements.

\begin{enumerate}[leftmargin=6mm]

    \item[\textbf{R1}] \textbf{Adaptive task planning:} Task proposals need to be tailored to users' analysis goals and the given dataset.
    This includes exploring in depth and breadth, with automatic planning and execution reducing user efforts.
    Moreover, as exploration results emerge, new tasks should contextually link to previous \revised{exploration results}, adapting to the switching between tasks. (C1)

    \item[\textbf{R2}] \textbf{Flexible visualization generation:}
    The generation of visualizations should flexibly handle different tasks and data.
    This encompasses accurate data identification, transformation, and selection of visualization types, as well as adding highlights based on discovered insights to reduce cognitive load. Furthermore, generated views should support interaction, allowing for more immersive user analysis. (C2)

    \item[\textbf{R3}] \textbf{Automatic insight generation:} To reduce development costs, the system should efficiently complete code-based data analyses for given tasks,  providing visualizations and suggesting findings to facilitate hypothesis forming and validation. To lessen the cognitive burden, important parts of insights should be highlighted in rich text format in the output results. (C2) 

    \item[\textbf{R4}] \textbf{Multiple view composition:} As new visualizations are added, older ones may become less relevant. To display the most recent results to the user, the visualization panel needs to be updated. However, it is crucial to avoid discarding previous results as they may be relevant to new insights. Users should be allowed to merge visualizations of interest even if they are not the latest. Within the merged views, users can focus their analysis on a smaller scope through interaction, reducing cognitive load among large sets. (C2)
        
    \item[\textbf{R5}] \textbf{Intuitive analysis process:} \revised{The relationship between tasks, visualization, insights, and analysis progress should be presented in a more intuitive form. The system should support interactions between human and agent to help users understand the agent's task planning and execution. (C1, C2)}

\end{enumerate}

\vspace{-10pt}
\subsection{Conceptual Framework}
\label{sec:flow-definition}

\begin{figure}[t]
    \centering
    \includegraphics[width=\linewidth]{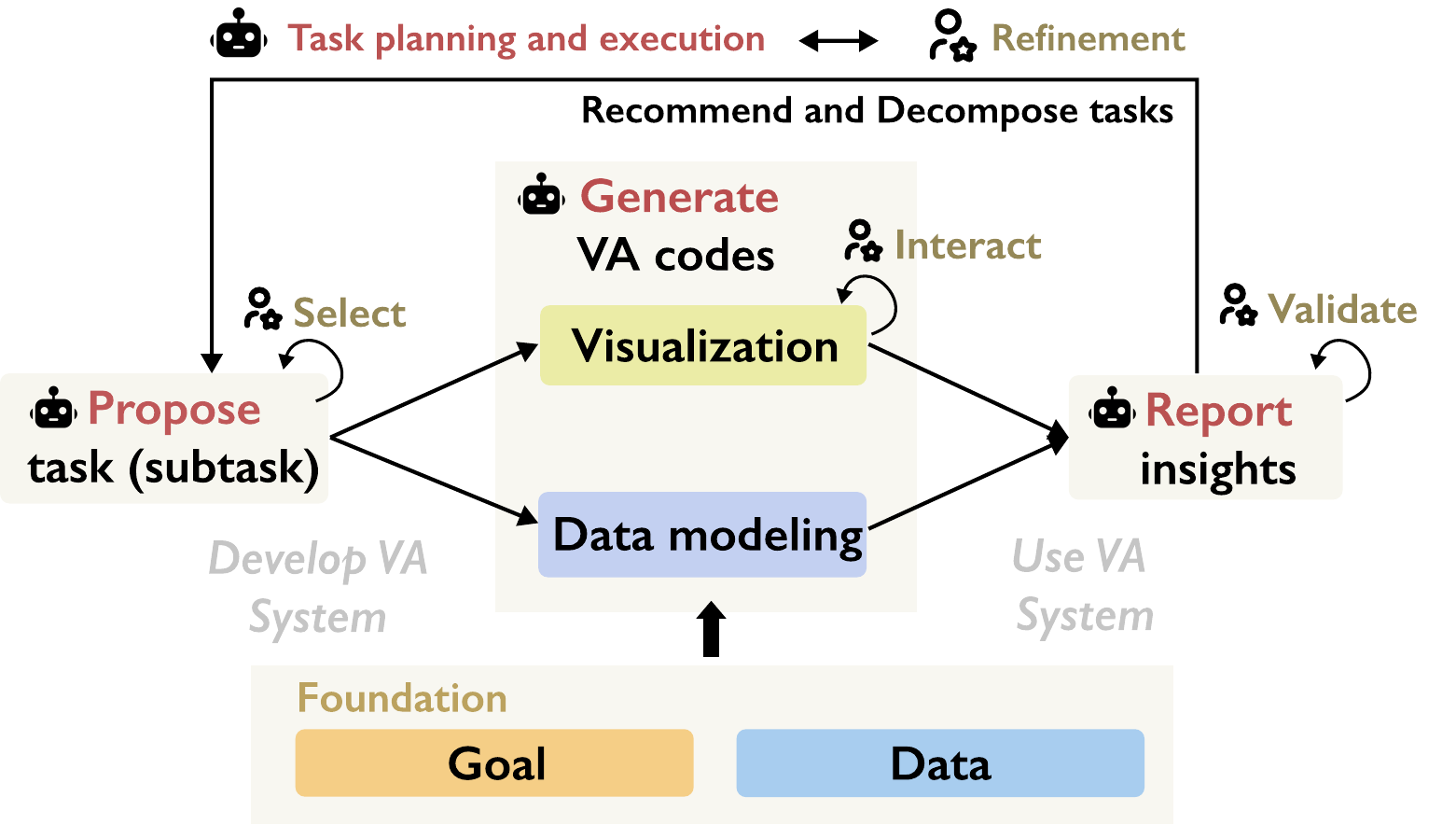}
    \vspace{-0.5cm}
    \caption{
        The illustration of conceptual framework in LightVA. 
        \rerevised{The} agent creates the VA system for analysis by task planning and execution, while the user uses the created VA system and refines agent outputs.
    }
    \label{fig:workflow}
    \vspace{-0.5cm}
\end{figure}

Based on the challenges and requirements discussed, we propose LightVA, a lightweight VA framework with agent-based task planning.
The ``lightweight" refers to the light expectations in developing VA systems and using the developed systems for analysis.
We design the framework to involve a recursive task-solving process in which goals and data are inputs, and insights are outputs.
The intermediate results are tasks, subtasks, visualization, data modeling~(\Cref{fig:workflow}).
The LLM agents are integrated to support goal understanding and task decomposition (R1), data modeling and visualization codes generation (R2, R3), and linked view generation for interactive exploration (R4).
\rerevised{Meanwhile, the users can monitor the process, guide the agent, and refine the agent's output through direct manipulations and natural languages (R5).}

\subsubsection{Defining Primitives}
\label{sec:primitives}

According to the above descriptions of the framework, some key primitives need to be defined.

\textbf{Goal and Data:} \revised{We define a goal as an overall description or a vague utterance that expresses the high-level purpose of the analysis.} For instance, \textit{``to analyze the influence factor on a car's fuel efficiency''.} \revised{The goal directs the focus of the analysis, and the data supplies the analysis evidence to meet the goal.}

\textbf{Task:} \revised{Tasks are specific aspects derived from the goal and make the \textit{goal} executable. We define them using four attributes:

\small
\begin{equation}
task := \langle type,~data~variables,~method,~progress \rangle
\end{equation}
\normalsize 

The \textit{type} indicates the nature or category of the analytical operation to be performed, such as finding extreme, outlier, change point, trend analysis, etc. The \textit{data variables} are the objects upon which the tasks will be applied.
The \textit{method} defines how the task will be solved, including data modeling and visualization methods, and the solved result is \textit{insight}.
If the task is complex, which means it covers multiple \textit{data variables} and needs to use multiple \textit{methods} to solve, the task can be decomposed into \textit{subtasks}.
To depict whether the task is completed, we define \textit{progress}, which means the task is solved when its subtasks are solved as well.
Given that the tasks can be complex and range from low to high abstraction, we describe them using natural language.
}

\textbf{Insight:} This denotes the expected or targeted knowledge or understanding that the task aims to achieve. 
We define insight using four attributes:

\small
\begin{equation}
insight := \langle type,~parameters,~data~variables,~data~values \rangle
\end{equation}
\normalsize

The \textit{insight type} can be discoveries, patterns, trends, or anomalies identified through analysis~\cite{Ding2019QuickInsights}. \revised{For example, if the task type is ``compare'', the insight type could be ``difference''~\cite{Wang2020DataShot}.} The \textit{parameters} specific features of the insight, such as ``increasing'' or ``decreasing'' of  ``trend''.
\revised{The \textit{data variables} for insights can be the original data columns or transformed variables, while the \textit{data variables} for tasks are the original data columns. For example, if the task involves analyzing vehicle weight and fuel efficiency (MPG), the task's \textit{data variables} would include ``Weight\_in\_lbs'' and ``MPG''.
If the insight includes a derived metric such as ``MPG per pound,'' this would be considered a transformed variable specific to the insight.
The \textit{data values} refer to particular values of \textit{data variables}. For example, the maximum MPG value is ``48''.}

\textbf{Visualization:} Refers to the visual representation required for the task that best conveys the data and insights.

\small
\begin{equation}
visualization := \langle type,~encoding,~interaction,~coordination \rangle
\end{equation}
\normalsize

\revised{Our framework generates visualizations using Vega-Lite grammar~\cite{satyanarayan2016vega}, which supports the above four aspects.} The types of \textit{interactions} are, e.g., filtering, zooming, and hovering supported by the visualization.
The \textit{coordination} describes how this visualization interacts or synchronizes with other visualizations, such as a brush, to filter each other.

%% file: sections/3.1-workflow-framework.tex
\subsubsection{Workflow of Framework}
\label{sec:framework-workflow}

\begin{figure}[t]
    \centering
    \includegraphics[width=\linewidth]{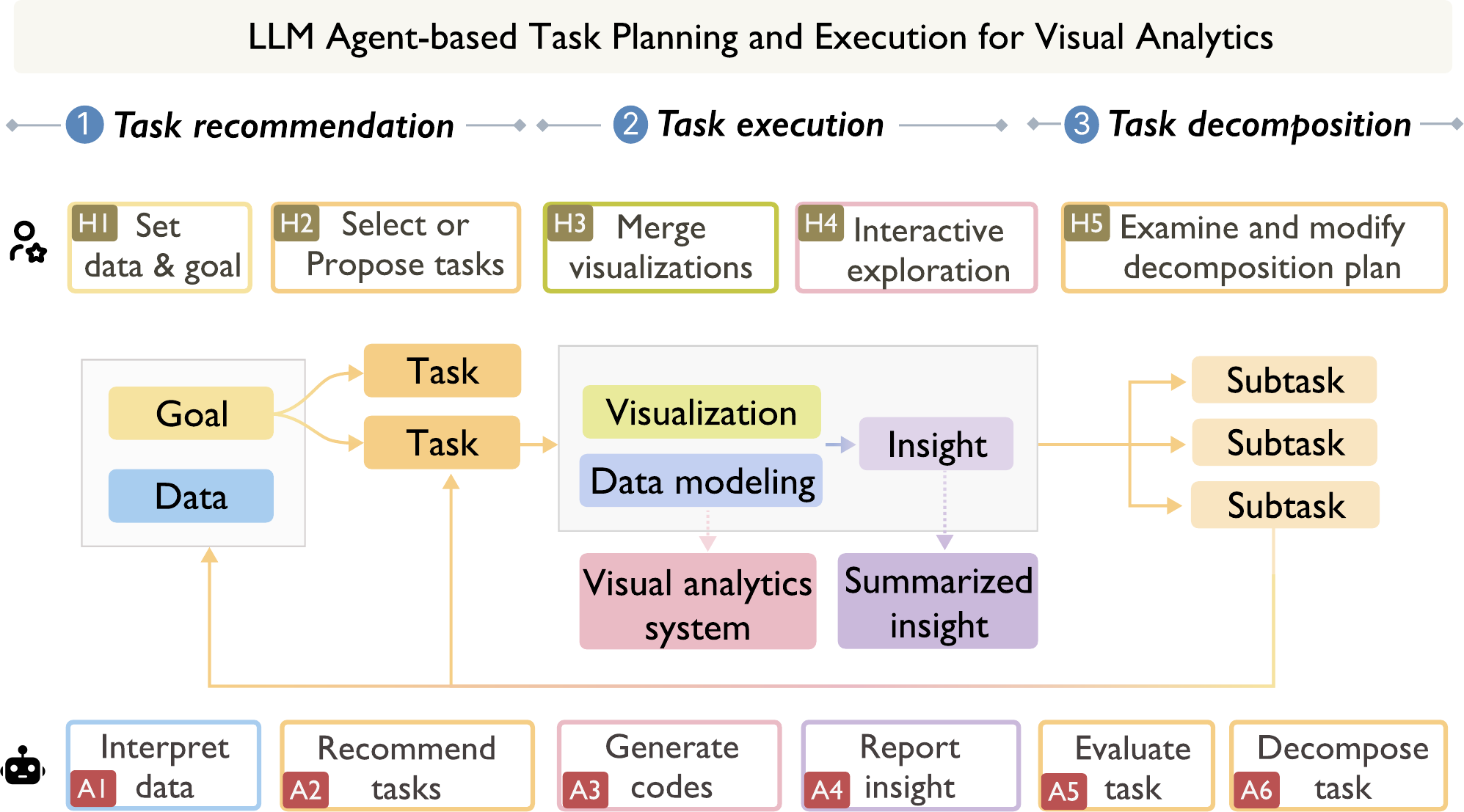}
    \vspace{-0.5cm}
    \caption{
        The workflow of the LLM agent-based task planning and execution for visual analytics.
        The collaboration between users and the AI agent is characterized by three stages: task recommendation, task execution, and task decomposition.
        The user proposes goals, selects tasks, merges visualizations, and engages in interactive exploration (H1-H5).
        The agent interprets data, recommends tasks, generates codes, reports insights, evaluates tasks, and decomposes tasks (A1-A6).
        }
    \label{fig:guidelindes}
    \vspace{-0.5cm}
\end{figure}

Our framework involves building a \textbf{task flow}, represented as a directed compound graph $G = (V, E)$, where each \textit{node} $v \in V$ represents a \textit{task} and \textit{edge} $e \in E$ represents the connection from one task to another task.
Since the process of exploration involves both enlightening and in-depth thinking, we define two types of task edges: \textbf{recommendation} and \textbf{decomposition}. And we define the solving of the task as \textbf{execution}.
The difference between recommendation and decomposition is that recommendation is ``goal-oriented'', and decomposition is ``task-specific''. The recommendation aims to broaden the exploration scope, providing heuristic suggestions. The decomposition aims to ensure a detailed plan with clear logic for solving the task. The following sections will introduce how humans and agents collaborated to complete the analysis in our framework~(\Cref{fig:guidelindes}).

\textbf{Stage1: Task recommendation.}
Initially, the user uploads the data and inputs a goal (\human{H1}), and then the agent interprets data (\agent{A1}) and transforms the goal into specific, actionable tasks (\agent{A2}).
The user can provide feedback and accept or modify recommended tasks to align with their analysis needs (\human{H2}).
Specifically, the interaction process involves two stages:

\begin{itemize}[leftmargin=*]
    \item \textbf{Initial stage:} At the begining, agent should first interpret the data and propose tasks that make the goal executable by mapping it to the data. The principle is to identify \textit{data variables} and 
    \textit{task type}. For instance, for a goal of \textit{``find high-risk events in a city''}, useful data variables could include time, space, text, and sentiment. The agent also needs to distinguish the purpose of the analysis, e.g., \textit{``finding outliers"}. Thus, different combinations of data variables and task types can form different tasks under a goal.
    
    \item \textbf{Historical context stage:} As the analysis continues to accumulate in the recommendation process, the agent should recommend tasks considering previous tasks and the overall goal. First, when a task is completed, the agent evaluates if the overall goal has been achieved. If not, it should recommend new tasks aligned with the goal. Second, previously unexplored tasks should be re-evaluated and proposed again if they are still relevant to the goal.
\end{itemize}

\textbf{Stage2: Task execution.} In this stage, the agent generates visualization and modeling codes (\agent{A3}) and executes the codes to report and summarize insights (\agent{A4}). Having multiple visualizations and modeling approaches, users could select visualizations to merge a linked view (\human{H3}) and interactively explore it (\human{H4}). It comprises the following two-step approach:

\begin{itemize}[leftmargin=*]
    \item \textbf{Visualization and insight generation}: When the user selects an agent-proposed task or proposes a task themselves, the agent writes codes to complete the analysis. The agent needs to choose appropriate data modeling and visualization \textit{methods} for each \textit{task} and run the codes to complete the analysis. The agent then needs to generate the \textit{insights} into a structured format. Finally, the agent should summarize insights obtained from the decomposition of subtasks, providing a comprehensive overview of the analysis. 
    \item \textbf{Multi-view linking}: Users can initiate \textit{coordination} by selecting multiple visualizations they prefer while the agent generates codes. We currently do not automatically combine visualizations from subtasks because subtasks decomposed from a main task often share common variables and visualization types. For example, several subtasks might use latitude and longitude to create maps exploring different spatial patterns. The linked view is often used to conduct multi-variate association analysis with different visualizations. Thus, allowing users to choose their visualizations can provide a more flexible analysis.
\end{itemize}

\textbf{Stage3: Task decomposition.}
We design the decomposition following a \textit{``on-demand''} strategy. In other words, execution is prioritized, and decomposition is considered only if the task is not completed. This approach aims to ensure users can see initial results quickly in an interactive environment. In this stage, the agent is responsible for evaluating tasks (\agent{A5}) and proposing a decomposition plan (\agent{A6}), while the users can examine and modify agent output to override the agents (\human{H5}). 

\begin{itemize}[leftmargin=*]
    \item \textbf{Results assessment}: Based on the initial execution's insights and the complexity of the task, the agent assesses the need for further analysis. According to the definition of \textit{task}, the agent should verify the selection of \textit{data variables} and the rationality of data modeling and visualization \textit{methods}. The evaluation should be explained to make users understand the motivation of decomposition. 
    \item \textbf{Sub-task generation}: If decomposition is necessary, agent should formulate a plan outlining subtasks. Each subtask should have appropriate \textit{data variables} and \textit{methods} addressing distinct aspects of the main task. These subtasks should have an execution order, e.g., in parallel or sequentially.
\end{itemize}

%% file: sections/4-flow-model.tex
\section{LightVA System}

Guided by the design requirements, we propose a pipeline of LLM agent-based task planning and an interface to support interactive visual data exploration with assistance.

\begin{figure*}[ht]
    \centering
    \includegraphics[width=\linewidth]{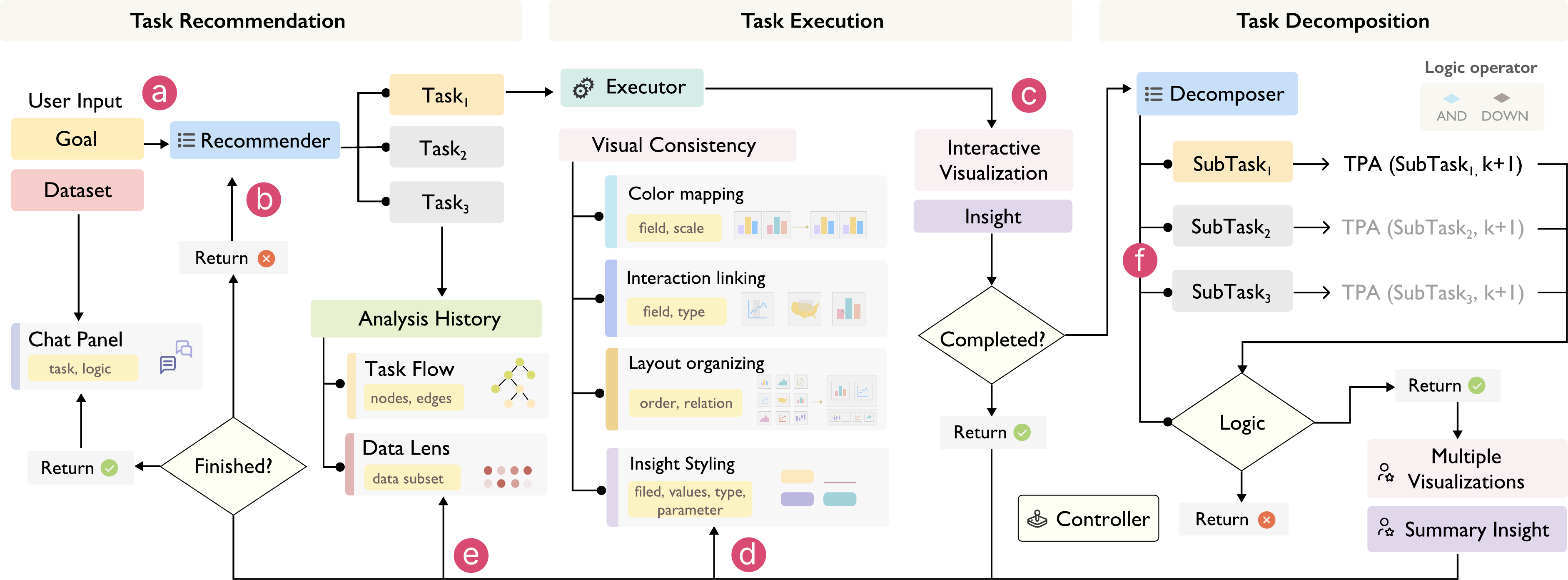}
    \vspace{-0.5cm}
    \caption{
        The agent-based system architecture.
        The pipeline starts from the goal and data with an agent-based task planning strategy, including recommendation, execution, and decomposition.
        \revised{Specific components and interactions are labeled (a)-(f), where (a) is the initial stage to recommend tasks from the goal, (b) recommends tasks with historical contexts, and (c) is the execution of tasks generating visualization and insights. After that, a series of optimizations for visual consistency are performed (d). Historical records are managed, and the progress of tasks is updated in a timely manner (e). According to the analysis results, the agent evaluates whether the task needs to be decomposed (f).}
    }
    \label{fig:framework}
    \vspace{-0.5cm}

\end{figure*}

\subsection{Agent-based Task Planning}
\label{sec:planning_process}

Based on the conceptual framework previously introduced, we propose an agent-based task-planning pipeline~\Cref{fig:framework}.
In this process, we have three kinds of modules. The \textit{planner} has two types: \textit{recommender} and \textit{decomposer}, which are responsible for task recommendation and decomposition respectively. The \textit{executor} handles task execution, including visualization generation and data analysis, and the \textit{controller} is the recursive algorithm that bridges the \textit{executor} and \textit{planner} (See the algorithm in the appendix). \rerevised{When introducing each stage's strategy, we provide the prompt templates that describe input and output, instructions to LLMs, and indicators to describe the output format.}
\rerevised{The prompt examples and outputs are available in Appendix A.}

\subsubsection{Task recommendation}
\label{sec:recommendation}

Task planning begins with a recommendation approach based on the given goal and dataset. The \textit{recommender} aims to transform the goal into actionable tasks. The implementation follows the step-by-step guidelines with two stages: initial stage and historical context stage (Prompt template~\ref{prompt1}). 

\revised{In the initial stage, the input is the goal and the data~(\Cref{fig:framework}-a) and the objective is to convert the goal into actionable tasks. 
\revised{Based on the framework (\Cref{sec:framework-workflow}), we guide the model with three principles. First, identify \textit{data variables} relevant to the \textit{goal}. Second, propose several \textit{task type} to refine the goal from different aspects. Third, draft descriptions of tasks by combining these data variables and task types.}

In the historical context stage (\Cref{fig:framework}-b), the generated tasks should be divided into two types: linking goals with discoveries to propose \textit{new} tasks and tasks that might have been forgotten by the user needs to \textit{review}.
To achieve this goal, the model might need to consider the following steps. First, evaluate completed tasks to check if the overall goal is achieved.
Second, propose new tasks based on the current context and previous tasks.
Third, propose previously unexplored tasks if they remain relevant.
In addition to guidelines, we should provide examples of output to guide models. } 

\begin{Prompt}[Task Recommendation]
    \textcolor{mycolor1}{----------------------------------------------------------\textit{Initial stage}}\\
    \textbf{Input}: \{\schemaname{goal, data}\}\\
    \textbf{Instruction}: You need to come up with a short plan based on the understanding of the data to help accomplish the goal. Please recommend n exploratory tasks, including task description, type, and data variables. \revised{For task type, you may consider the trend, correlation, category, distribution, etc, to explore the goal from different aspects. For data\_variables, list the original column names.\\
    \textbf{Indicator}: \{\schemaname{An example in JSON format}\}}\\
    \textbf{Output}: \{\schemaname{new tasks\}}\\
    \textcolor{mycolor1}{---------------------------------------------------\textit{Historical context}}\\
    \textbf{Input}: \{\schemaname{goal, data, explored and unexplored tasks\}}\\
    \textbf{Instruction}: You need to supplement some new tasks for explored tasks by considering the results of tasks already explored if needed. Second, it is recommended that previous unexplored tasks be revisited if they are suitable for analysis at this stage by considering the explored tasks.\\
    \textbf{Indicator}: \{\schemaname{An example in JSON format}\}\\
    \textbf{Output}: \{\schemaname{new tasks, \revised{existing tasks to review}\}}
    \label{prompt1}
\end{Prompt}

\subsubsection{Task Execution} 
\label{sec:execuetion}

After tasks are proposed and confirmed by the user, the executor will solve tasks by generating codes (\Cref{fig:framework}-c).
We then adopt the ``decompose on-demand'' strategy, which means we first execute the task and then decompose the task if the results are unsatisfactory. 
This approach allows users to obtain preliminary results quickly and enables a more flexible way to address both simple and complex tasks, avoiding unnecessary time cost that comes from always decomposing tasks in advance.

To execute a task, the input for the \textit{executor} includes a goal, data, and task description, and the output includes visualizations and insights for the selected task (Prompt template~\ref{prompt2}).
In the first round, two code snippets are generated: one for data analysis and the other for visualization. In the second round, structured insights are generated by data analysis results.

\begin{Prompt}[Task Execution]
    \textcolor{mycolor1}{----------------------------------------------------\textit{Code generation}}\\
    \textbf{Input}: \{\schemaname{task, data summary, code template}\}\\
    \textbf{Instruction}: You need to write Python codes to analyze the data to solve this task. After finishing the data analysis, you should continue to use Altair to generate an interactive visualization. Add a brush or click function, a tooltip, and a legend if different colors are used.\\
    \textbf{Indicator}: \{\schemaname{A code template}\}
    \begin{lstlisting}[language=python]
import altair as alt
import pandas as pd
def plot(data: pd.DataFrame):
    # Data preprocessing
        <codes>
    # Chart generation
    chart = alt.Chart().mark_bar().encode()
    return chart
    \end{lstlisting}
    \textbf{Output}:\{\schemaname{insight, visualization}\}\\
    \textcolor{mycolor1}{-------------------------------------\textit{Structured insight generation}}\\
    \textbf{Input}: \{\schemaname{task, data, codes}\}\\
    \textbf{Instruction}: \revised{Run the \schemaname{codes} to report an important insight for this task: \schemaname{task}. 
    You should output insight, including text, insight type, parameters, data variables, and data values. \{\schemaname{Definitions of insight attributes.}\}}\\
    \textbf{Indicator}: \{\schemaname{A few examples in JSON format}\}\\
    \textbf{Output}: \{\schemaname{insight}\}
    \label{prompt2}
\end{Prompt}

\textbf{Visualization generation:} For each task, the model needs to implement data analysis and visualization methods, where one task corresponds to one visualization, and the visualizations will be combined into multiple views when multi-variables association analysis is required, which we will introduce later.
We use Vega-Lite via Altair~\cite{satyanarayan2016vega} to generate visualizations. Vega-Lite is a high-level grammar that can support the generation of a variety of visualization types in a low-code way, and its declarative structure allows users to modify visualizations easily. 
As Altair only supports basic data transformation, we leverage Python’s flexible libraries, allowing for more complex analysis, such as regression and clustering. However, this may introduce inconsistencies and errors between analysis and visualization code snippets. To address this, we provide a code scaffold in the prompt to let LLMs fill the empty to enhance consistency and improve output stability.

\textbf{Structured insight generation:} 
After generating visualization and modeling codes, the executor needs to execute them to structure the results into insights. 
As the agent's coding environment\footnote{\url{https://platform.openai.com/docs/assistants/tools/code-interpreter}, Last accessed March 2024. Code Interpreter allows Assistants to write and run Python code in a sandboxed execution environment.} does not support Altair, we use the local environment to execute the visualization part of the code. The agent runs the data analysis part, interprets the results, and derives insights. 
We provide a format example according to our definition of insight in the framework to guide model output. This includes the insight description, the insight type, and the data variables and values.
\revised{These attributes highlight key elements in the insight description to improve readability.}
    
\textbf{Linked-view generation}: For multi-variable association analysis, we enable the generation of multiple linked views (\Cref{fig:framework}-d). While LLMs can add interactivity and set basic colors and layouts within the code, maintaining visual consistency across views, such as avoiding duplicate colors and ensuring neat layouts, requires implementing additional constraints.

\begin{itemize}[leftmargin=*]
    \item \textbf{Interaction linking:} To enable interaction linking among charts, we instruct the model to modify and combine the chart codes following several guidelines. First, ensure both charts contain common key columns with consistent data formats for the selection fields. Next, define the selection mechanisms. Common interaction methods include a brush on the time axis for the line chart, dual brushes for the scatter plot, and click interactions for the bar chart. Then, the transform filter should be applied to the other charts.
    \item \textbf{Layout organization:} When generating new visualizations, we set the charts to the same size and arrange them sequentially. If the user selects charts to merge multiple views, we allow selecting up to six charts to avoid excessive cognitive load. The LLMs should output the layout no more than three charts in a row.
    \item \textbf{Color mapping:} We built a data-to-color mapping rule within the task space based on \rerevised{Qu and Hullman’s} guidelines \cite{qu2017keeping}, where one color corresponds to one data dimension without reuse. For example, the same field should use the same quantitative color scale across different views, while different fields should use non-overlapping hues or palettes to avoid confusion.
\end{itemize}

\textbf{Task progress calculation: }
In the analysis process, each node has a \textit{progress} attribute that quantifies the task completion degree. When the task is executed, the agent evaluates the task's completion status. If the task remains incomplete and needs further in-depth analysis, the progress is set to 0\%; otherwise, it is set to 100\%. 
Upon determining a task as incomplete, the agent suggests a decomposition plan. 
If the further decomposition is denied by the user, the progress is set to 100\%. 
Each recently executed task is represented as a leaf node, initiating a bottom-up refresh of progress values for non-leaf nodes across the tree~(\Cref{fig:framework}-e).
The process for the main task is the average progress of two child tasks, which are user-confirmed nodes generated by task decomposition or recommendation. 
This hierarchical update process accurately reflects the analysis status and ensures progress values account for both the agent's assessments and user input.

\begin{figure*}[t]
    \centering
    \includegraphics[width=\linewidth]{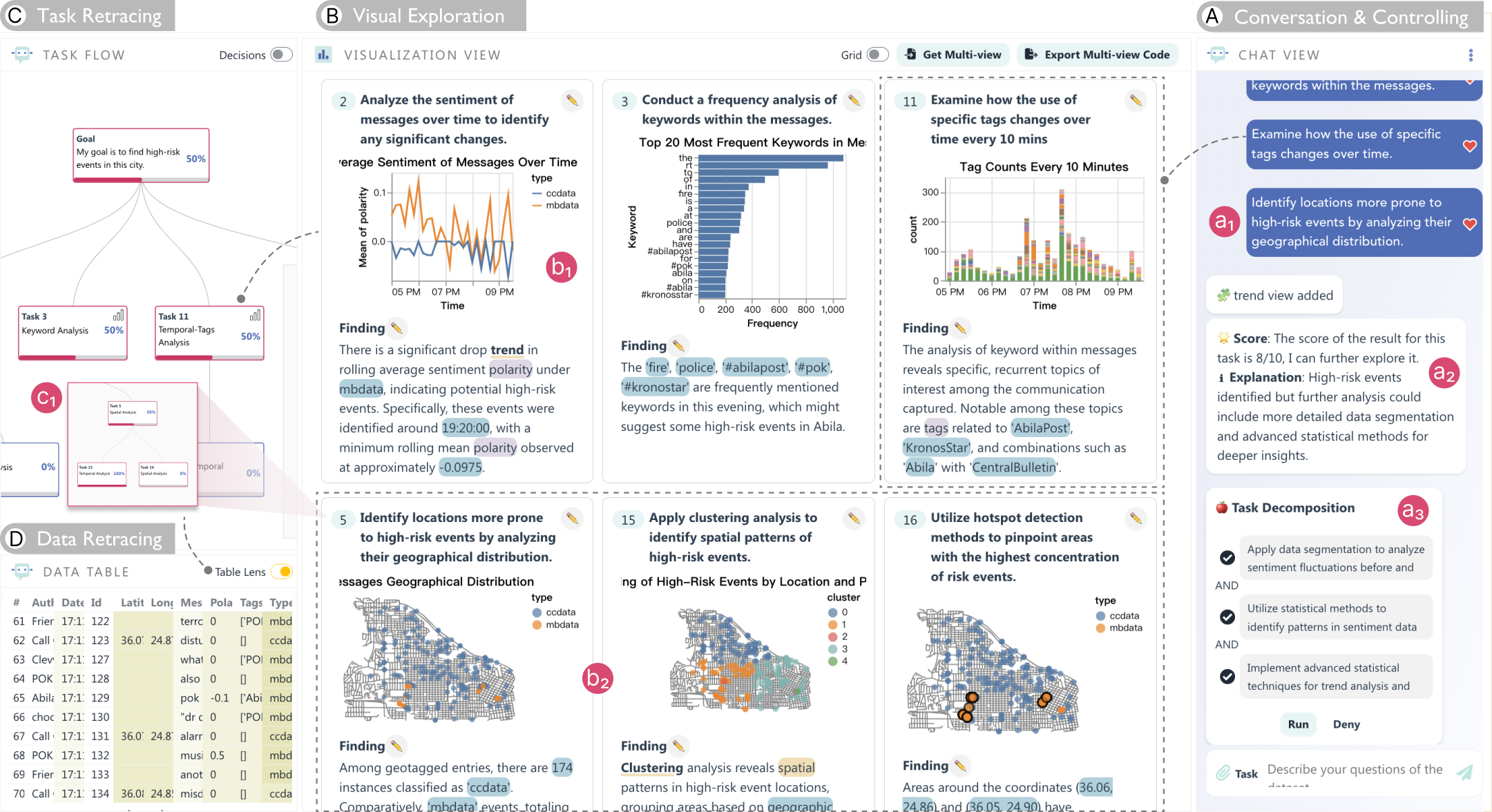}
    \vspace{-0.5cm}
    \caption{The LightVA system comprises four views. Users can communicate with LLMs and control the planning process in (A) Chat view by selecting the tasks or setting the decomposition plan. The generated visualization and insights from LLMs are updated in (B) Visualization view. Task flow view (C) visualizes the task planning structure and allows users to control the analysis process. When a task is completed, users check the data exploration situation in the Data table with table lens (D). }
    \label{fig:system}
    \vspace{-0.5cm}
\end{figure*}

\subsubsection{Task Decomposition} 
\label{sec:decomposition}

Upon the execution of a task, the \textit{decomposer} evaluates the quality of completion and decides whether to decompose and how (Prompt template~\ref{prompt3}). To evaluate if the task is completed, the input includes tasks, codes, and insight, while the output is a score ranging from 0 to 10 and an explanation. Then, if decomposition is needed (i.e., its score is below a certain threshold, like 8), a detailed decomposition plan with the appropriate logic operators will be proposed~(\Cref{fig:framework}-f). Conversely, if no further decomposition is required, the \textit{controller} calls the \textit{recommender} to propose new exploratory tasks. \revised{According to the framework, the evaluation process can be guided by the following guidelines. The first is to determine if the initial solution adequately segments the \textit{data}, ensuring that the analysis covers all relevant subsets.
Second, assess if the task requires further data mining or visualization \textit{methods} (e.g., regression, clustering) to extract deeper insights.}

\begin{Prompt}[Task Decomposition]
    \textcolor{mycolor1}{-----------------------------------------------------------\textit{Is complete}}\\
    \textbf{Input}: \{\schemaname{task, codes, insight}\}\\
    \textbf{Instruction}: You need to judge whether the task requires further analysis. The complexity of the task can be considered from data, data mining, and visualization methods. If the task appears incomplete, it needs to be decomposed further. Please rate the task from 1-10. For example, if the initial solution adequately segments the data or if the task requires advanced statistical analysis. You should output a score and explanation for your evaluation.\\
    \textbf{Indicator}: \{\schemaname{A output template in JSON format}\}\\
    \textbf{Output}: \{\schemaname{score, explanation}\}\\
    \textcolor{mycolor1}{-----------------------------------------------------------\textit{Decompose}}\\
    \textbf{Input}: \{\schemaname{task, insight, score, explanation}\}\\
    \textbf{Instruction}: This task requires further analysis. According to the score and explanation, please generate no more than n subtasks and indicate the methods they use respectively based on this. There are two operators AND and DOWN in the execution order to connect tasks.\\
    \textbf{Indicator}: \{\schemaname{An example in JSON format}\}\\
    \textbf{Output}: \{\schemaname{subtasks, execution order}\}    
    \label{prompt3}
\end{Prompt}

If a task requires decomposition, a depth-first decomposition process is employed. \revised{This process is guided by two factors. One is a predefined maximum decomposition step depth ($k_{max}$), managing the depth ($k$) of decomposition (decompose when $k<k_{max}$).} 
Another factor is the completeness of subtasks. Here, we define each task that can be decomposed into subtasks with two types of logic. 

\begin{itemize}[leftmargin=*]
    \item \textit{AND}: Subtasks are independent and executed in parallel order with multiple agents.
    \item \textit{DOWN}: This is a particular \textit{AND} case. Subtasks are dependent and will be executed in serial order with a single agent.
\end{itemize}

\revised{We use an example to illustrate the decomposition process. If the task of \textit{analyzing vehicle weight and fuel efficiency} is completed with a score of 6/10, the agent might suggest the following decompositions:
T1: \textit{Segment the data into weight categories (light, medium, heavy) and analyze the fuel efficiency within each segment.} 
T2: \textit{Within each weight category, conduct a clustering analysis to identify patterns or groupings that could further explain variations in fuel efficiency.}
T3: \textit{Perform a multiple regression analysis to control for additional variables like engine size and vehicle age}.
The execution logic would be (T1 DOWN T2 AND T3).

According to the logic, the agent executes each subtask and summarizes the subtasks' insights to formulate an overall insight for the decomposed task.
No further decomposition is required if subtasks are all completed or up to the max steps.
Then, considering the goal and analysis results, the agent might propose revisiting previously proposed tasks or recommending new exploration tasks.
}

%% file: sections/5-framework.tex
\subsection{LightVA Interface}
\label{sec:interface}

The agent-based interface includes several views to enable user-controlled visual exploration, as shown in~\Cref{fig:system}.
To enable this, we provide four views: \textit{Chat view}, \textit{Visualization view}, \textit{Task flow view}, \textit{Data view}.
The design of the interaction follows the design considerations of \Cref{sec:requirements} and provides three modes of interaction.

\textbf{Chat view:}
The exploration begins with the \textit{Chat view}~(\Cref{fig:system}A), an LLM-based chat box that facilitates direct communication between users and agents through natural language interaction. The interaction between the user and agent can switch between delegate, guide, and discuss.
In this interface, the user uploads data and inputs goals. The agent then responds with its understanding of the dataset and suggests new tasks in the form of buttons.
Users can bookmark tasks of interest, and tasks that are not of interest will not be counted in the progress of exploration. The user can then execute a task by clicking on it. Additionally, users can enter their own tasks in the chat box.
After submitting a task, the agent will provide a visualization of the production and insights obtained from the calculations in the \textit{Visualization view}~(\Cref{fig:system}B). If the task needs to be decomposed, the task decomposition plan will appear.
The user can modify the logical relationship between the task and the execution plan \revised{by switching the text button}.
The agent will analyze the task based on the user's instructions and return the results of multiple subtasks.
During the exploration process, users can ask various questions and discuss them with LLM in the dialog box, such as the problem analysis method, understanding of the answer, and explanation for the recommendation.

\textbf{Visualization view:}
This view presents visualizations and insights in the form of cards. Each card contains the task serial number, the task content, the interactive visualizations, and the insights in rich text format. Users can modify the Vega-Lite JSON codes and insight text. To address the issue of cognitive load for the user, we allow the user to select which visualization cards they want to view, and these selected cards are displayed in the main view while the remaining cards are moved to the candidate set below. Additionally, the user can merge the selected cards to create an interactive linked view. \revised{Moreover, users can modify and export the generated codes, which providing flexibility for those who need to customize their analyses further.}

\textbf{Task flow view:}
The task flow view updates as goals and tasks are added, providing the user with a clear analysis of status and progress~(\Cref{fig:system}C). Each node is a rectangular box showing the task id, task type, and percent progress. The original task text appears when the cursor is over a node. Clicking on a node will update the visualization layout in the \textit{Visualization view}. Hovering over the edge can see the differences in data and task type between the associated nodes. In accordance with design considerations, we allow users to choose unexplored tasks to be executed and delegate them to the agent. Additionally, users can remove pending tasks from the flow to reduce workload. A clear exploration structure may inspire user's ideas. 

\textbf{Data table view:}
In addition to task visualization, we provide a data lens visualization in \textit{Data table view}~(\Cref{fig:system}D) to guide users in the large exploration space. When a task is being completed, the user can observe the exploration of the accumulated data usage frequency in table lens mode. The color of the table cell represents the relative frequency of each cell being explored. This observation may inspire the user to discover regions of interest to propose new questions.

\revised{\subsection{Error Handling}

We refer to \textit{errors} as issues that cause the system to crash or become unresponsive.
Due to the inherent unpredictability of LLM outputs, errors may occur if the outputs cannot be parsed correctly, causing the system to crash or become unresponsive.
To ensure the system operates correctly and prevents workflow disruptions, we implement an error-handling mechanism.

To develop an effective error-handling mechanism, we conducted a test using two datasets in expert evaluation with 20 agent-opposed tasks, employing both GPT-3.5-turbo and GPT-4-turbo.
Each task was tested with code generation and insight annotation, utilizing two different prompting techniques.
This resulted in a total of 160 initial tests. 
We classified the errors in the test into five categories: \textit{(1) Unfamiliar dataset, (2) Data binding issues, (3) Serialization issues, (4) Data transformation issues, (5) Syntax errors}. More details about the test can be found in Appendix B.
To address these types of errors effectively, we implemented specific error-handling strategies in three ways: before model generation with prompting techniques, within the system after generation, or delegated to users.

\textbf{Prompting techniques:} (1) Few-shot prompting~\cite{fewshot2020}: We provide examples to assist the model in better grasping the requirements of the outputs. For example, when making insight annotations, examples can explain the insight components of the model. (2) Chain-of-thoughts~\cite{wei2022chain}: We guide the model through a thought plan with step-by-step instructions, which is helpful in reasoning processes, such as task execution.

\textbf{Within system handling:} (1) Self-reflection~\cite{Shinn2024reflection}: We allow LLMs to examine and correct their actions and outputs. For example, the model should debug codes based on the observation of the error. Based on our test results, the self-correction helped reduce around 40\% initially identified errors, such as syntax errors, spelling mistakes, and logical inconsistencies. (2) Catching and feedback: Common syntax errors such as matching quotes and parentheses, can be solved by rule-based solutions. We classify these common errors to highlight which steps in the data analysis process the LLM made mistakes rather than merely pointing out low-level errors.

\textbf{User-side handling:}
Once errors are identified, the system notifies the user, and the user may edit the codes.
Additionally, we maintain the analysis history, allowing for a rollback to the previous step if an error occurs during the execution of the current task.
Users can choose a new task and remove the failed task from the task flow. 

}

%% file: sections/6-usage-scenario.tex
\section{Usage scenario: Event Analysis}

To examine the effectiveness of our framework, we demonstrate the LightVA with IEEE VAST Challenge 2021 Mini-Challenge 3~\footnote{\url{https://vast-challenge.github.io/2021/MC3.html}} as a usage scenario.
The challenge's goal is to detect events that happened in Abila City during the evening of January 23, 2014. The provided data include microblog records and emergency dispatch records from a call center. 
Our motivation for this scenario stems from two reasons.
Firstly, it incorporates a representative blend of data types and corresponding visualizations, encompassing text, spatial, and temporal data, which is a typical VA scenario.
Secondly, the VAST challenge has the ground truth to support an objective evaluation of our LightVA. We use the OpenAI GPT-4 model in our work.
\revised{
For a more detailed demonstration of this scenario, a video is added to the supplemental materials.}

\textbf{Initialization:} In the beginning, we upload a dataset .csv file and a map outline data .json file at the \textit{Chat view}.
Then, we type a goal, ``find some high-risk events in this city''.
The dataset is introduced briefly, and the system proposes four initial exploration tasks based on our goal: \textit{sentiment analysis, keyword analysis, tags analysis, and spatial analysis}~(\Cref{fig:system}-a1).
As the four tasks are aligned with our objectives, we bookmark them all in our progress.
We then prioritize the exploration of the first task - \textit{sentiment analysis}.
Upon submission, the agent generates a time varying card at the \textit{Visualization view}~(\Cref{fig:system}-B).
The generated insight for sentiment analysis indicates that the microblog records has experienced a drop in average sentiment, especially with a minimum polarity in 19:20, indicating potentially dangerous events~(\Cref{fig:system}-b1).

\begin{figure}[t]
    \centering
    \includegraphics[width=\linewidth]{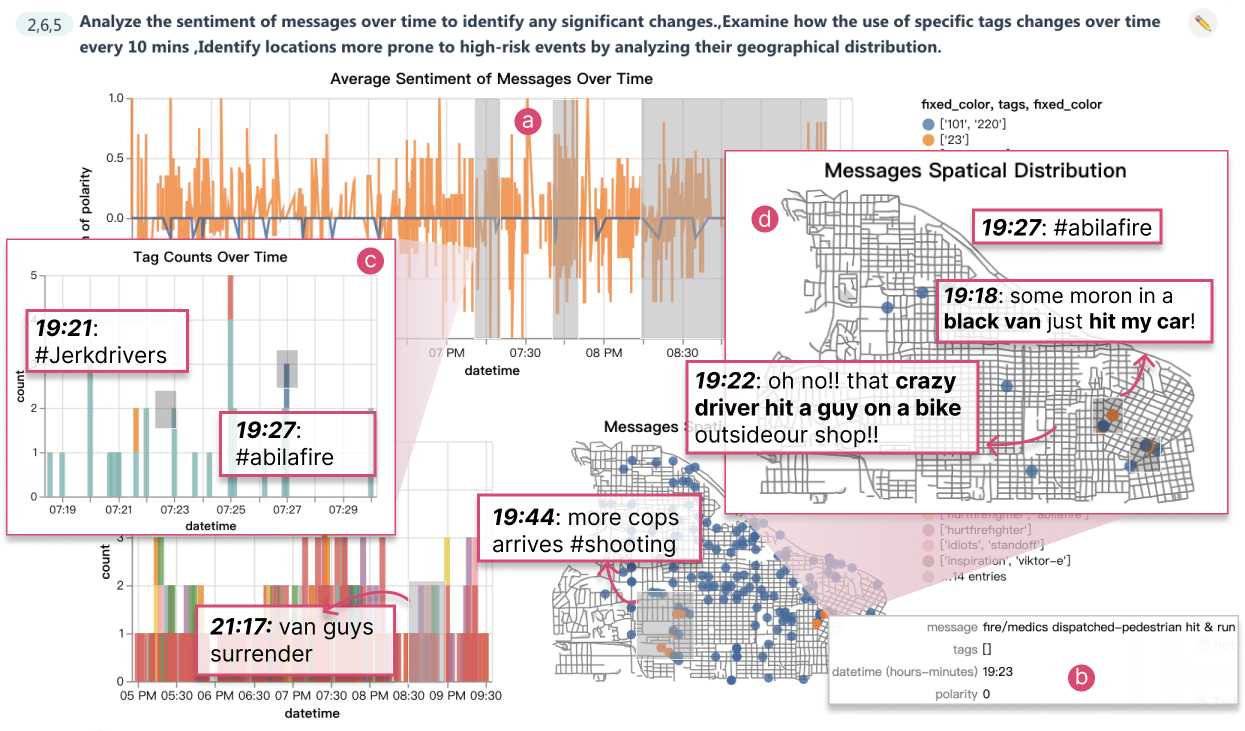}
    \vspace{-0.5cm}
    \caption{
        \revised{
        A linked view for event analysis.
        The timeline, histograms, and map are selected by the user to generate a linked view.
        }
        The charts can be brushed or clicked to filter each other (a).
        The polarity, tags, and message can be found in the tooltip (b) of each view to support detailed analysis.
        By interactively exploring the linked views (c, d), we find some dangerous events, such as ``Hit and run'', ``Standoff'', ``Fire''.
    }
    \label{fig:vast1}
    \vspace{-0.5cm}
\end{figure}

\textbf{Task recommendation:} The agent not only generates the visualization and findings but also evaluates them. For this \textit{sentiment analysis} task, the agent gives a score of 8/10 and explains that the generated results have basically completed the task, but further statistical analysis can be performed~(\Cref{fig:system}-a3). As we prefer to start with a broad overview, we decide not to decompose. The agent thus reminds us to revisit our previous collection tasks in time.
Thus, we choose the second task (\textit{keyword analysis}), which involves generating a preliminary keyword visualization. 
The generated finding summarizes the highly frequent words such as ``fire'', ``police'', and ``POK''.
In addition to the agent's finding, clicking over the stacked histogram for task 11 shows the time range of each word. 
For example, the ``shooting" from 19:40 until 19:50 and ``abilafire'' from 19:00 until 21:30. 
These findings suggest high-risk incidents like shootings and fires at first glance. The agent then recommends \textit{spatial analysis}. 
From the \textit{Data table} with lens, we can notice that there is indeed no latitude and longitude explored so far, which indicates that agent recommendation can pay attention to the data coverage~(\Cref{fig:system}-D).
After execution, the system generates a map with shapes and messages, and the decomposition structure is updated on the Task flow~(\Cref{fig:system}-c1). 
We observe a higher concentration of microblogs in two specific locations, which might indicate incidents affecting public safety in those areas~(\Cref{fig:system}-b2).

\textbf{Task decomposition:} Based on the assessment of tasks and preliminary results, the agent recommends further decomposition with three subtasks: \textit{clustering analysis, hotspots detection, and prediction modeling}, with an AND and DOWN logic~(\Cref{fig:system}-a3).
To conduct a retrospective analysis, we choose the first two subtasks.
The cluster analysis combines location and polarity to divide messages on the map into five categories. We find that two places on the left and right sides of the map have a distinct sentiment distribution. 
The second sub-task, through analyzing message quantity, highlights hotspots, visually indicating areas with dense messages. 

\textbf{Linked-view analysis:}
Based on the suggestion of the agent, we can form the analytical logic from multi-variables to concrete details.
\revised{To find events more clearly, we select three tasks from the task flow view. The tasks include sentiment evolution charts, tag evolution charts, and a map to form an interactive multi-view~(\Cref{fig:vast1}).}
Through interaction on line chart~(\Cref{fig:vast1}-a), we observe the histogram and map and discover that around 19:20, a ``hit-and-run'' event occurs: a ``black van'' first collides with a small car and then a cyclist, sparking a lot of discussions~(\Cref{fig:vast1}-b, c).
Additionally, we find that the fire occurred near the hit-and-run site.
Later, we find that the black van heads west by 19:44, and a shooting occurs with the police, marking the period with the highest discussion intensity.
Finally, at 21:17, the van guys surrender.
By observing the table and task flow, we find that the progress of the goal reached 100\%, and the data is basically covered, ``type'' ``latitude'' and ``longitude'' were explored frequently. 

To conclude, in this scenario, we find significant events such as Fire, Hit and Run, and Stand-off without manually coding and designing.
With task planning, we generate eight views with auto-summarized findings based on statistical analysis and a linked view to solve the goal rapidly.
Compared to our submission to the VAST Challenge~\cite{Peng2021Mixed}, where we spent 2 people, 2 days, and 6 hours on data preprocessing (including tagging, categorizing, and removing spam messages) plus 2 people, 7 days, and 6 hours on interface construction and analysis, totaling around 108 hours.
In LightVA, we upload the preprocessed data.
Further construction and analysis take roughly 1 hour.
This represents an efficiency increase of approximately 5 times, as the current effort is 25 out of 108 hours.
Moreover, it is worth noting that data wrangling still require human involvement, especially in complex scenarios.
One potential avenue would be to integrate agent-assisted data annotation and cleaning into LightVA's workflow.
In addition, there may be a lack of finesse and aesthetic appeal in interface visualizations and interactions when compared to manual designs.
Despite this, the overall cost-effectiveness has been enhanced.
To address this concern, we offer support for exporting the code, which can be further modified as needed.

%% file: sections/7-user-study.tex
\section{Expert Study}

To examine the effectiveness of LightVA in VA system construction and task recommendation, we invited two VA experts (denoted as E1 and E2) and a domain expert (denoted as E3) to participate in the study.
These experts have experience in data analysis and are familiar with using VA systems. 

\textbf{Participants background:} 
E1 is a VA expert specializing in VA system development, topics around digital humanities, text-based data, road data for autonomous driving, spatiotemporal datasets, and tabular data.
E1 often uses D3.js and Vue.js to develop VA systems.
E2 is a VA expert with experience in autonomous driving and social media, as well as work experience in business analysis. 
In daily work, E2 chooses to use tools like Tableau within the company to analyze quantitative data.
The frequency of using data analysis and visualization tools is about weekly.
E3 is a domain expert analyzing data for fast-moving consumer goods and supply chains. 
E3 frequently uses Excel and Power BI and occasionally uses Python for in-depth analysis. 
The use of data analysis and visualization tools occurs on a daily basis.

\textbf{Procedure:}
The study consists of three sessions.
First, we spent 15 minutes to know the experts' backgrounds in data analysis and introduced our work by showing the video demonstration.
Then, in the exploration phase, the experts spend about 30 minutes exploring the system using the think-aloud method~\cite{van1994think}. 
Considering E3's daily work requirements and aligning with their domain expertise, we opted for a dataset that mirrors their routine tasks.
Due to data confidentiality concerns, we substituted the original dataset with a comparable one related to sales for E3's use. 
Meanwhile, the automotive dataset was designated for exploration by E1 and E2, fitting their respective areas of expertise.
During the system usage stage, we observed and recorded how they interacted with the LightVA system.
In the final stage of the interview, we spent approximately 30 minutes gathering the experts' feedback on their usage experience of the system.

\subsection{Visual Analytics Expert Evaluation: Cars Dataset}
\label{sec:cars}

In this study, we evaluate LightVA with E1 and E2 using a well-known Auto MPG dataset~\footnote{\url{http://archive.ics.uci.edu/ml/datasets/Auto+MPG}}.
The dataset has 406 records and 9 attributes, including brand, model, performance indicators (such as horsepower and cylinders), and the year of manufacture and place of origin.

\begin{figure}[t]
    \centering
    \includegraphics[width=\linewidth]{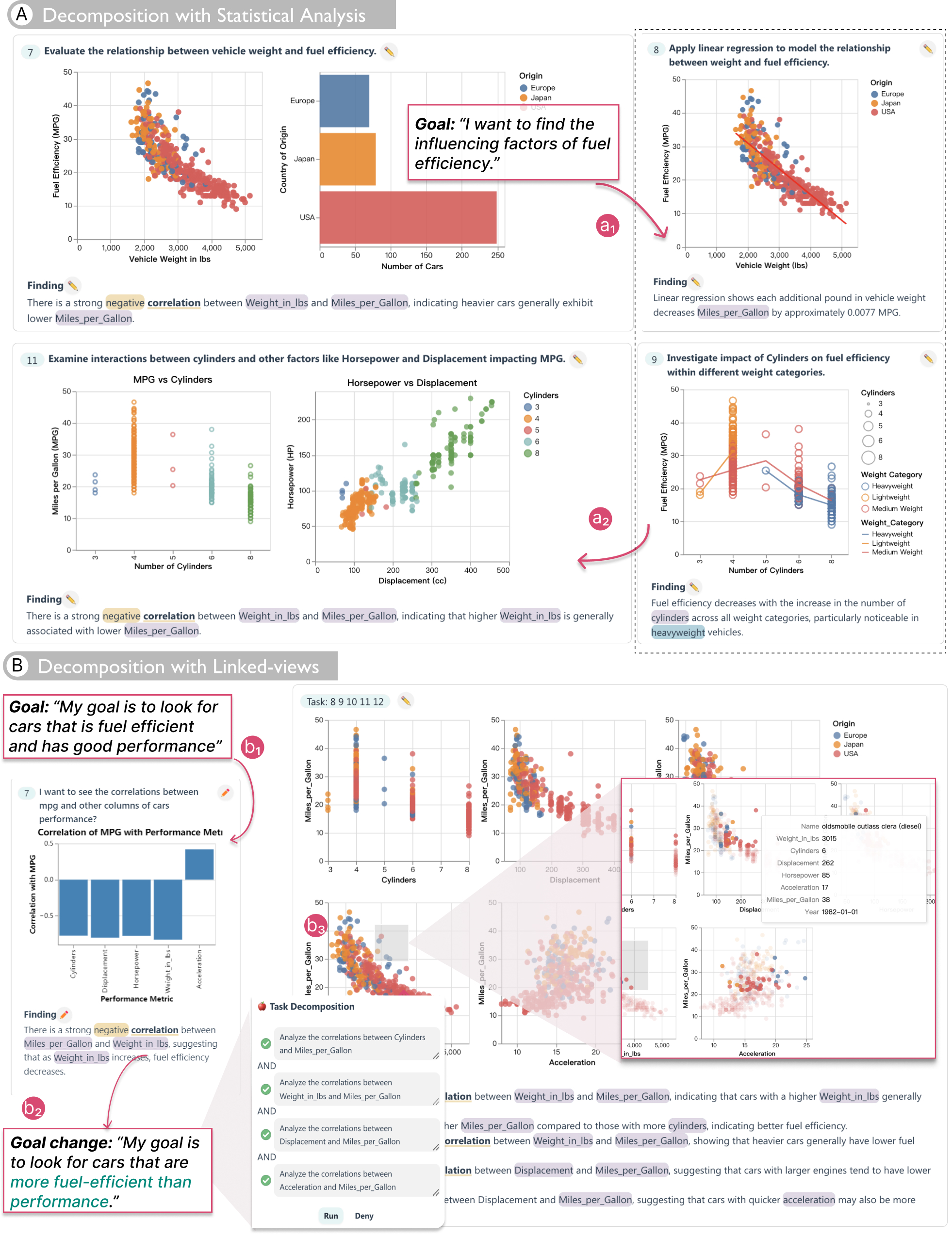}
    \vspace{-0.5cm}
    \caption{The illustrations of the expert study using our framework with cars dataset.
        Two VA experts show different analysis preferences during the exploration process.
        (A) E1 focuses on the creation of VA, progressively decomposing the goal and employing statistical methods to complete the exploration.
        (B) E2 has a clear goal during the analysis, and continuously optimizes the goal with task decomposition, and leverages the linked-view interactions to find a satisfactory answer.
    }
    \label{fig:cars-study}
    \vspace{-0.5cm}
\end{figure}

During the exploration process, E1 is more concerned with the construction of the VA system.
At the beginning, E1 sets the analytical goal to \textit{identify the factors influencing fuel efficiency}.
E1 chooses one of the initially proposed tasks for execution.
After completing that task, the agent presents a detailed decomposition plan.
E1 then follows the agent's guidance to continue selecting subsequent sub-tasks for further decomposition and in-depth analysis.
In the generated visualization (\Cref{fig:cars-study}-a1), E1 discovers the differences in the distribution of the number of vehicles produced by different origins across various performance ranges in the bar chart on the right, by utilizing the area brushing feature on the scatter plot.
E1 also discovers the performance differences between cars with different numbers of cylinders through the filter functionality between linked view (\Cref{fig:cars-study}-a2).

Another expert E2, demonstrates a focused interest in the data analysis, pursuing a clear analytical goal that was refined throughout the analysis process.
Initially, E2's goal was \textit{to identify vehicles that excel both in fuel efficiency and performance}~(\Cref{fig:cars-study}-b1).
However, recommended insights revealed a strong negative correlation between fuel efficiency (measured in MPG) and the majority of performance metrics, indicating a trade-off between high fuel efficiency and superior performance capabilities.
This insight led E2 to quickly adjust the analytical goal towards \textit{prioritizing fuel efficiency while ensuring performance was not significantly compromised}~(\Cref{fig:cars-study}-b2).
Based on the plan proposed in task decomposition, E2 explored the relationship between various vehicle performance metrics and fuel efficiency with five charts.
Finally, E2 combined these charts to generate linked view~(\Cref{fig:cars-study}-b3).
By filtering points across views, E2 found several vehicle models meeting the revised goal.

\subsection{Domain Expert Evaluation: Sales Dataset}

To further evaluate LightVA, we conducted another study with a domain expert.
E3 was interested in a sales dataset for a large store that provides detailed transaction information~\footnote{\url{https://www.kaggle.com/datasets/addhyay/superstore-dataset}}.
The dataset contains 3,312 records with 21 fields, e.g., sales, discount, profit, and category.

At the beginning, E3 mentioned the high-time sensitivity of sales data for products in real-world scenarios.
Therefore, in the initial tasks, E3 opted to explore the temporal trends of product sales volumes.
In the histogram and insight (\Cref{fig:sales-study}-a1) suggested by the agent, E3 found that the overall sales of the item differ significantly from month to month.
In the next step of the decomposition plan, E3 categorized the products to explore the temporal trends of different categories (\Cref{fig:sales-study}-a2).
It was found that the temporal trends of the different categories were broadly similar, but the total sales volume of the technology category in November was the highest.
Moreover, following the agent's decomposition into another task: \textit{``Analyze sales trends of top-selling and bottom-selling products''}, E3 noted that in the domain scenario, \textit{``finding out how well products are selling can optimize inventory to prevent stockouts and excess inventory.''}
Therefore, E3 selected a task previously proposed but not executed, recommended for review by the agent, to rank the sales volumes of different categories and subcategories (\Cref{fig:sales-study}-a3).
Based on the generated visualization, E3 discovered that the ``phone'' subcategory within the ``Technology'' category has the best sales performance.

After finding the answer to the previous task, E3 was interested in \textit{``analyze the correlation between sales and profit.''}
Upon receiving the specific scatter plot (\Cref{fig:sales-study}-b1), E3 discovered that high sales do not always mean higher profit.
The expert speculated that this may be due to the impact of discounts, as items with high sales often have significant discounts.
E3 confirmed this hypothesis through interaction with and analysis of the scatter plots of sales and profit, with and without discount (\Cref{fig:sales-study}-b2).
To find the optimal discount rate for products, E3 chose to analyze the relationship between discount and profit across different product categories.
According to \Cref{fig:sales-study}-b3, the differences between product categories are minor, with all reaching maximum profit at a 10\% discount.

%% file: sections/7.1-results.tex
\subsection{Results and Analysis}

This section discusses the expert's exploration process and feedback on the system construction and task planning.

\textbf{Exploration process analysis:}
The exploration behaviors of the three experts differ significantly.
VA expert E1 created the most visualization charts, with half including interactive features, emphasizing the generation of visualizations and the exploratory use of linked views.
Furthermore, E1 delved deeper into task decomposition to employ more complex statistical methods, such as linear regression and polynomial regression.
E2, with profound data analysis experience, did not strictly follow the agent's recommendations during the exploration process.
Instead, E2 refined and improved the analysis goals and exploration directions based on the results of previous tasks, proposing new questions and flexibly employing interactive VA to complete the goal.
E3, as a domain expert, possessed an understanding of the characteristics of certain data attributes within the dataset, such as the relationships between discounts, sales, and profit.
E3's analysis was grounded in real-world scenarios, selecting different analysis methods as needed to integrate data insights into practical applications.
This suggests our system supports a degree of flexibility, responding effectively to users' individual needs.

\textbf{Feedback on system construction:} E2 and E3 both expressed that interactive analysis through linked views facilitates a better understanding of the data.
From the perspective of a VA expert, E1 commented that \textit{``basic charts and interactions can be generated, which are sufficient for simple data analysis problems, but modifications might be necessary for more complex issues.''} 
E2 and E3 agreed that the system can effectively generate insights, enhancing the accuracy of analysis. 
\revised{E2 suggested that generating insights based on awaring of user behavior would enhance the execution of tasks.}
Additionally, E3 noted the significant reduction in manual effort due to AI generation, stating, \textit{``Previously, using PowerBI required a lot of time and operations, but now it only takes a sentence.''
}

\textbf{Feedback on task planning:}
E3 observed that \textit{``the proposed tasks are quite good, indicating the system has a certain understanding of the dataset, and the language used is very standard.''} 
E1 valued the system's ability to decompose tasks as \textit{``the most useful part,''} which can provide deeper analysis and reveal certain characteristics of the data, especially for those lacking domain-specific expertise.
However, domain expert E3 cautioned that the decomposed tasks are not always reliable, stating, ``The process of task decomposition needs to be explained.''
\rerevised{Regarding task recommendations, E1 seeks more targeted suggestions that adhere closely to instructions without becoming too divergent. In contrast, E3 emphasizes that approaching goals from different perspectives can yield valuable insights and strategies in their work scenarios. These differing viewpoints highlight the need for the planning algorithm to be adaptable to user preferences.}

\begin{figure}[t]
    \centering
    \includegraphics[width=\linewidth]{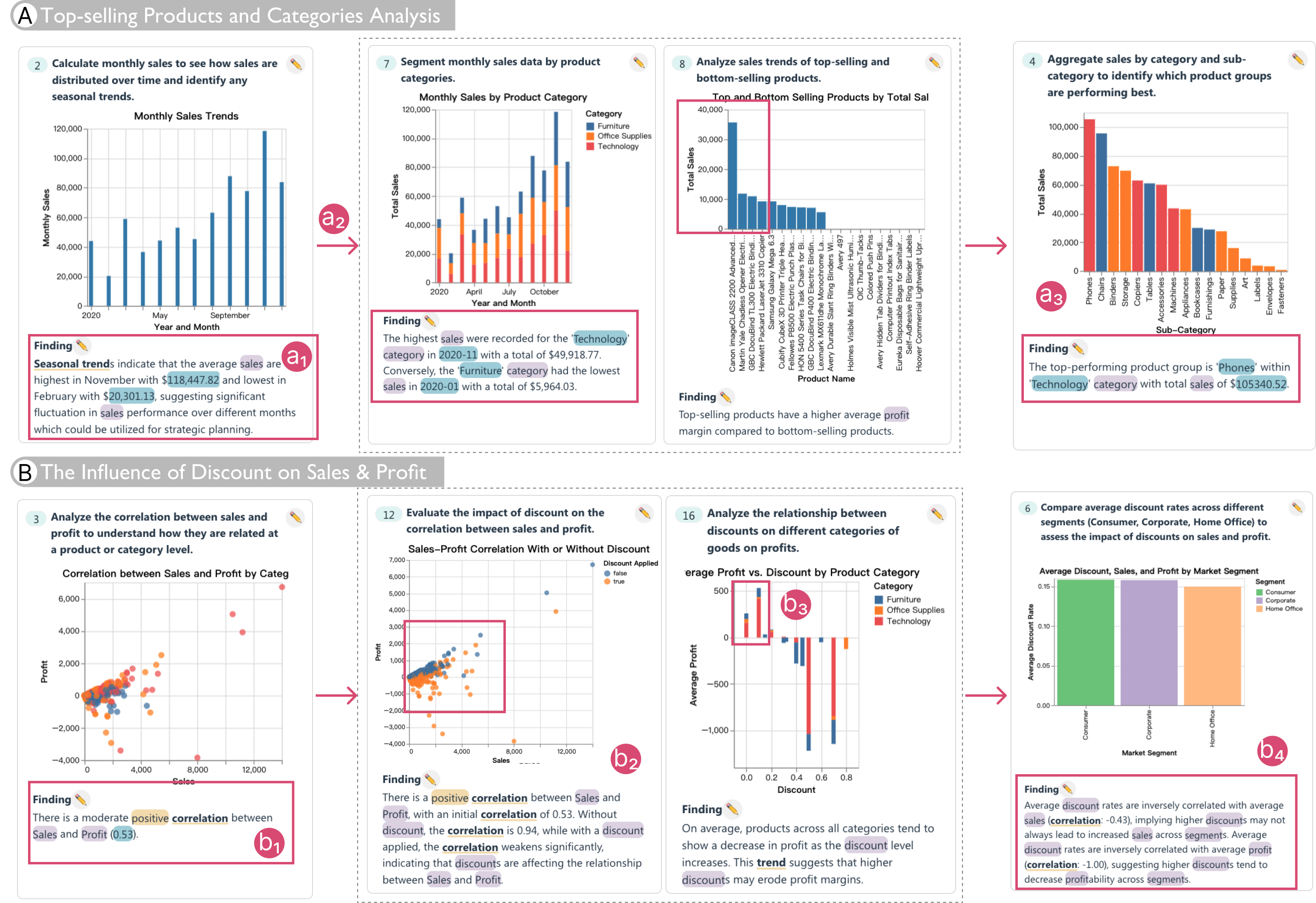}
    \vspace{-0.5cm}
    \caption{The example results of the domain expert study with a superstore sales dataset. E3 implements analysis from overview to details and beyond with two rounds of task decomposition. (A) The top-selling products and specific categories are found for inventory formulation. (B) The system effectively pointed out the impact of discounts on sales and profits, promoting hypothesis generation and verification.}
    \label{fig:sales-study}
    \vspace{-0.5cm}
\end{figure}

%% file: sections/8-discussion.tex
\section{Discussion}
\label{sec:discussion}
\revised{This section outlines the limitations of our work and discuss the implications learned from the research with the future directions.}

\revised{\textbf{Generalizability of the framework:}}
Our framework is generalizable in two aspects. 
First, our conceptual framework unifies the development and usage of VA systems, based on the connections between goals, tasks, visualizations, and insights. 
Secondly, compared with existing visualization software, e.g., Tableau (with AI version, Ask Data)~\footnote{\url{https://help.tableau.com/current/pro/desktop/en-us/ask_data.htm}, last accessed March 2024}, LightVA offers several distinct advantages. 
While Tableau supports natural language interfaces, task recommendation, and single visualization generation, it lacks support for task decomposition, multi-view visualization generation, and iterative human-in-the-loop task completion evaluation, which are the focuses of LightVA.

Meanwhile, there are some limitations to our framework. First, the framework may require more detailed guidelines for complex and specific domains, particularly in task customization, visualization methods, and model selection. A more comprehensive grammar for tasks and insights would improve model output evaluation and refinement. This necessitates extensive research and documentation, categorizing different tasks and domains to enhance the framework's applicability. 
\rerevised{Second, functionality in the development of VA systems needs to be improved, particularly features that facilitate authoring and version iteration.}
Third, the relatively small number of participants in the expert study limits the generalizability of the results. Conducting a larger crowdsourced user study would be beneficial in verifying the quality of the automated output further and investigating the factors that contribute to users' perceived quality.

\revised{\textbf{The performance of LLMs in VA tasks:}}
While LLM exhibits potentials, some of the limitations of LLM requires careful handling.
\begin{itemize}[leftmargin=*]
    \item \textbf{Output stability and accuracy}:
    The output of LLMs can be unstable.
    During our evaluation, we encountered instances where the model failed to follow instructions accurately, leading to parsing and execution failures in the generated code.
    \revised{In order to address this problem, we propose an error-handling mechanism.
    Although this mechanism helps to avoid errors making system crashes, additional LLM errors about incorrect facts should be detected and corrected.
    In the future, we can incorporate LLMs' self-reflection to solve hallucination~\cite{ji2023towards} and provide insights on errors so that users can work together to fix these errors}.
    
    \item \textbf{Response speed}:
    \revised{From our test, GPT-3.5-turbo completes tasks in significantly less time than GPT-4-turbo.
    To increase the running speed, we use a multi-agent parallel computing strategy and error handling for time exceeding a certain threshold.}
    A future direction could be to utilize caching mechanisms such as GPTCache~\cite{bang2023GPTcache} to explore acceleration strategies in data analysis scenarios.
    
    \item \textbf{Problem-solving ability}:
    \revised{Our test results indicate that LLMs can struggle with complex tasks, such as predictive modeling.
    This highlights the need for a broader evaluation of model capabilities across various complex domain-specific problems to derive guidelines or evaluation metrics for task planning and execution.
    Additionally, we found that LLMs may struggle with processing issues like missing values due to unfamiliarity with data.
    From the implementation side,} we should provide additional materials or tools and teach agents to process data effectively~\cite{Qin2023Toolllm}.
    \item \textbf{Domain knowledge}:
    In some scenarios, LLMs may not have sufficient domain knowledge.
    To address this limitation, \revised{future research can focus on combining retrieval augmented generation (RAG) and fine-tuning for LLMs to solve specific domain tasks}~\cite{GaoFineTuned}.
\end{itemize}

\textbf{Injecting visualization design knowledge: }
Design knowledge is important in designing VA systems.
In our implementation, we constrain color, interaction, and layout.
These constraints are not exhaustive and we only consider them as preliminary.
Further research can include more design guidelines for the LLM with prompting techniques or multimodal models~\cite{ZengTuning} to perceive and evaluate the effectiveness.
Besides, a memory module~\cite{packer2023memGPT} can be integrated to make the agent evolve.
In addition, interactions other than natural language can be designed to facilitate user input of intentions, such as sketching~\cite{DirectGPT} or generating widgets~\cite{Chenglong2024DynaVis}.

\textbf{Automation and personalization:}
From the user study, we found that users from different backgrounds have different needs. Some users have a clear idea of what they want and need the agent to follow orders in detail, while others prefer the agent to take the lead in a more automated way.
This divergence in user preferences highlights the balance between user agency (the control they maintain over decisions) and the automation of processes. In future studies, we could explore how different levels of agency and automation affect task performance and user satisfaction~\cite{lin2020dziban}.
Additionally, users may require different levels of assistance in terms of breadth and depth. We could conduct further qualitative experiments to test the differences in decision-making paths when experts in visualization are assisted by agents~\cite{ battle2019characterizing, Liu2020Paths}, comparing these paths visually using graph algorithms. This could help us build a preference knowledge base for different user types, enabling the model to simulate and adapt to users' desired paths, providing more personalized recommendations.

%% file: sections/9-conclusion.tex
\section{Conclusion}

In this paper, we aim to reduce the complexities and technical demands of carrying out visual analytics.
Therefore, We introduce LightVA, a \revised{lightweight visual analytics framework that supports task planning, insight analysis, and linked visualization generation based on human-agent collaboration}. Our framework utilizes LLM agents for task planning and execution.
\revised{The framework employs a recursive approach in which the agents recommend tasks, break down complex tasks into subtasks, and generate visualization and data modeling codes to solve tasks.
We develop a system that embodies our proposed framework, supporting users to analyze data based on the communication with LLM agents and use the task-driven generated VA system.}
A usage scenario and an expert study suggest that LightVA not only reduced the manual effort required but also provided new opportunities to leverage LLMs to facilitate visual data exploration.

%% file: sections/appendix.tex
\section{Implementation Details}
\label{appsec:appendix}

This appendix introduces prompt examples and a history of analysis during the analysis process.

\subsection{Task Planning and Execution Algorithm }

Here is the task planning and execution algorithm we introduce in~\Cref{sec:planning_process}.

\begin{algorithm}[h]
\caption{ \fullform{}}
\begin{algorithmic}[1]
\Function{\method{}}{Goal $G$, Task $T$, Data $D$, History $H$, Decomposition\_depth $k = 1$}
    \If{T is empty}
        \State $T_{\mathrm{new}} \gets \boldsymbol{\mathrm{Recommender}}(G, D)$
            \State $H \gets \boldsymbol{\mathrm{Update}}(H,T_{\mathrm{new}})$\Comment{Add new tasks to history}
            \State \Return $T_{\mathrm{new}}, H$
    \Else
        \If{$k > k_{\text{max}}$} \Return $\{\}, False$ \EndIf
        \State $result \gets \boldsymbol{\mathrm{Executor}}(T, D)$ 
        \State $completed \gets \boldsymbol{\mathrm{Decomposer}}(T, result)$
        \Comment{Evaluate if the task is completed and output score and explanation}
        \If{$completed$ is False}
            \State $subtasks, logic \gets \boldsymbol{\mathrm{Decomposer}}(T, D)$
            \For{each $T_\mathrm{sub}$ in $subtasks$}
                \State $result[T_\mathrm{sub}], H \gets \method{}(G, T_\mathrm{sub}, D, H, k + 1)$
            \EndFor
            \State $completed \gets \mathrm{logic}(result)$
            \State \Return $completed, H$
        \Else
            \State $T_{\mathrm{new}} \gets \boldsymbol{\mathrm{Recommender}}(G, T, D, H)$
            \State $H \gets \boldsymbol{\mathrm{Update}}(H,T_{\mathrm{new}})$ \Comment{Add new tasks to history and update progress of tasks and goal}
            \State \Return $T_{\mathrm{new}}, H$ 
        \EndIf
        \State \Return $completed, H$
    \EndIf
\EndFunction
\end{algorithmic}
\label{algorithm}
\end{algorithm}

The algorithm aims to achieve a specified goal through recursive task decomposition and execution. Given a Goal \( G \), Task \( T \), Data \( D \), History \( H \), and Decomposition\_depth \( k \), the algorithm initially checks if the task is empty. If so, it calls a task recommender to generate a new list of tasks, which is then updated in the history. If a task \( T \) is already defined, the algorithm checks whether the current decomposition depth exceeds a maximum depth \( k_{\text{max}} \); if it does, it returns an empty result and a failure flag. 

If the depth is within limits, the algorithm executes the current task using an executor and then assesses task completion with a decomposer. If the task is incomplete, the decomposer further breaks down the current task into a list of subtasks, which are then recursively processed by the algorithm. The results are combined based on a logical structure to determine overall task completion. The algorithm updates the history with any newly recommended tasks and finalizes completion status accordingly.

\subsection{Prompt Examples}
\label{appsec:appendix_prompt}

This section provides examples of prompts and output results when using LLM in multiple stages, including task recommendation, visualization generation, insight generation, is completed, decompose planner, merge visualization and summarize insights.

Our prompt template introduced in \Cref{sec:planning_process} consists of four abstract components:

\begin{itemize}[leftmargin=*]
    \item \textit{Input}: The data or context provided to the model that is necessary for executing the task. Example: A summary of the data, the task description, and a code template.
    \item \textit{Instruction}: Detailed directions that guide the model on how to perform the task. Example: Instructions to write Python code for data analysis and visualization using Altair, including adding interactive features like brush functions, tooltips, and legends.
    \item \textit{Indicator}: Examples or formats that illustrate what the expected output should look like. Example: A partial code snippet or formatted output that helps the model understand the structure and format of the final result.
    \item \textit{Output}: The result generated by the model based on the input and instructions provided. Example: The generated Python code, the resulting visualization, and a JSON object summarizing the insights.
\end{itemize}

\subsubsection{Task Recommendation}
This stage's prompt is designed to recommend new and previously unexplored tasks based on the goal and the analysis history. 

\begin{Prompt}[Propose initial tasks]
    Your need to come up with a short plan to help a user accomplish the goal:\{goal\}. Please recommend \{n\_0\} exploratory tasks. For task type, you may consider the trend, correlation, category, distribution, etc., to explore the goal from different aspects. Do not use advanced modeling methods. At the end of your answer, you should write the data summary and tasks beginning with \textasciigrave\textasciigrave\textasciigrave json and ending with \textasciigrave\textasciigrave\textasciigrave,. Summary no more than 100 words. Each task is a sentence. For data\_variables, the task may not correspond to the data, so you need to go to the dataset and find the corresponding data\_variables, and even if transformation is required, you need to write the original column name. There may be more than one column of data so it's an array. No need to say Here is the requested JSON format listing the tasks:. Do not add other sentences after this json data.
    \begin{lstlisting}[language=json]
{
     "tasks": ["","",""],
     "data_variables":[],
     "task_type": ["","",""]
}
    \end{lstlisting}
\end{Prompt}

\begin{Prompt}[Propose new tasks]
    Based on the goal: ``My goal is to analyze the fuel efficiency", previously explored tasks and unexplored tasks, \\
    (Extracted from analysis history)\\
    please recommend 3 new tasks. The dataset named cars.csv has already been uploaded for you.\\
    First, supplement some new tasks for explored tasks by considering the results of tasks already explored if needed. You just need to record the id in source in your json output, if no just leave it none. Every new task has a source.\\
    Second, it is recommended that previous unexplored tasks be revisited if they are suitable for analysis at this stage by considering the explored tasks. You just need to record the id in tasks to review in your json output, if no just leave it none.\\ You also need to label a task type for new tasks.\\ 
    At the end of your answer, you should write the tasks beginning with \textasciigrave\textasciigrave\textasciigrave json and ending with \textasciigrave\textasciigrave\textasciigrave, and the json data includes a list of new tasks and their source, the reviewed tasks and insight type will need to be analyzed. Like this:
    \begin{lstlisting}[language=json]
{
    "tasks": ["","",""], 
     "source": ["","",""],
     "tasks_to_review":[],
     "data_variables":[],
     "task_type": ["","",""]
}
    \end{lstlisting}
    Each task is a sentence.
\end{Prompt}

An output example of recommendation.

\textasciigrave\textasciigrave\textasciigrave json
\begin{lstlisting}[language=json]
{
    "tasks": [
        "Determine the optimal number of cylinders that balance fuel efficiency and vehicle weight.",
        "Investigate the impact of year of manufacture on the relationship between weight and fuel efficiency.",
        "Perform a cluster analysis to identify groups of cars with similar weight and fuel efficiency profiles."
    ],
    "source": [
        "9",
        "3",
        "3"
    ],
    "tasks_to_review": [
        "4",
        "5"
    ],
    "task_type": [
        "Optimization",
        "Historical Analysis",
        "Cluster Analysis"
    ],
    "data_variables": [
        ["cylinders", "MPG", "Weight\_in\_lbs"],
        ["year", "Weight\_in\_lbs", "MPG"],
        ["Weight\_in\_lbs", "MPG"]
    ]
}

\end{lstlisting}
\textasciigrave\textasciigrave\textasciigrave

\subsubsection{Visualization Generation}
The following prompt is the visualization generation part of task execution. By inputting the task and data introduction, the agent generates the visual and data analysis code using Python.
\begin{Prompt}[Visualization Generation]
    For the given task ``Evaluate the relationship between vehicle weight and fuel efficiency.", data summary,\\
    (Add a data introduction)
    you need to write Python codes to analyze the data and solve this task. The dataset cars.csv has been read into a dataframe by pandas.read\_csv(). Just use the variable `data' to refer to this dataframe and do not read another data. After finishing the insight analysis, you should continue to use Altair to generate the visualization. Ensure that each generated visualization has a brush function that allows you to select a subset of the data. Please ensure that only one view is generated. No combination of views is required. Please give me the whole code, including data analysis and Altair, in a function called `plot(data)'. You only need to give the function definition of plot(data), not execute it.\\
    please write the codes in this structure:
\textasciigrave\textasciigrave\textasciigrave python \textasciigrave\textasciigrave\textasciigrave, there is an example: \\
\textasciigrave\textasciigrave\textasciigrave python
    \begin{lstlisting}[language=python]
import altair as alt
import pandas as pd

def plot(data: pd.DataFrame):
    # Data preprocessing
        <codes>
    # Chart generation
    chart = alt.Chart().mark_bar().encode().properties()
    return chart
    \end{lstlisting}
    \textasciigrave\textasciigrave\textasciigrave
\end{Prompt}

An output example of visualization generation.

\textasciigrave\textasciigrave\textasciigrave python
\begin{lstlisting}[language=python]
import altair as alt
import pandas as pd

def plot(data: pd.DataFrame):
    # Ensure correct data types
    data = data.astype({
        'Weight_in_lbs': float,
        'Miles_per_Gallon': float
    }).dropna(subset=['Weight_in_lbs', 'Miles_per_Gallon'])  # Remove rows with missing values

    # Create selection features
    click = alt.selection_multi(fields=['Name'])
    brush = alt.selection(type='interval', encodings=['x'])

    # Base chart for scatter plot
    scatter = alt.Chart(data).mark_circle(size=60).encode(
        x=alt.X('Weight_in_lbs', axis=alt.Axis(title='Vehicle Weight in lbs')),
        y=alt.Y('Miles_per_Gallon', axis=alt.Axis(title='Fuel Efficiency (MPG)')),
        color=alt.condition(brush, 'Origin:N', alt.value('lightgray')),
        opacity=alt.condition(click, alt.value(0.9), alt.value(0.2)),
        tooltip=['Name:N', 'Weight_in_lbs', 'Miles_per_Gallon', 'Year:T', 'Origin:N']
    ).add_selection(
        brush
    ).add_selection(
        click
    ).properties(
        width=600,
        height=400
    )

    # Base chart for histogram
    histogram = alt.Chart(data).mark_bar().encode(
        x=alt.X('count()', axis=alt.Axis(title='Number of Cars')),
        y=alt.Y('Origin:N', axis=alt.Axis(title='Country of Origin')),
        color=alt.Color('Origin:N', legend=alt.Legend(title="Origin")),
        opacity=alt.condition(click, alt.value(0.9), alt.value(0.2)),
    ).transform_filter(
        brush
    ).add_selection(
        click
    ).properties(
        width=600
    )

    return alt.vconcat(scatter, histogram)
\end{lstlisting}
\textasciigrave\textasciigrave\textasciigrave

In actual analysis, challenges may arise, such as overly large datasets, unclear meanings of data attribute names, or highly domain-specific data, which may prevent an LLM from fully understanding the uploaded dataset. Summarizing the dataset in natural language and using this as a prompt can help the LLM perform the analysis task more effectively.

The data introduction includes a natural language description of the column names of each variable in the data, the types of variables and some statistical indicators. The given output json format template and some existing descriptions of the data (or not) are passed to LLM as prompts and let it generate data introduction according to the given format.

This is an example that we provide for the cars dataset.
\begin{lstlisting}[language=TypeScript]
class DataProperty {
  column: string; // Column Name
  description: string; // Natural language description
  dtype: string; // Data type of the column
  std: number | null; // Standard deviation, null if not applicable
  samples: string[]; // Sample values from the column
  nunique: number; // Number of unique values in the column
  min: string; // Minimum value in the column
  max: string; // Maximum value in the column
}
\end{lstlisting}

\begin{lstlisting}[language=Json]
{
    "columns": [
        "Name",
        "Miles_per_Gallon",
        "Cylinders",
        "Displacement",
        "Horsepower",
        "Weight_in_lbs",
        "Acceleration",
        "Year",
        "Origin"
    ],
    "properties": [
        {
            "column": "Name",
            "description": "The make and model of the car.",
            "properties": {
                "dtype": "object",
                "std": null,
                "samples": [
                    "ford galaxie 500",
                    "oldsmobile delta 88 royale",
                    "ford mustang ii"
                ],
                "nunique": 311,
                "min": "amc ambassador brougham",
                "max": "vw rabbit custom"
            }
        },
        {
            "column": "Miles_per_Gallon",
            "description": "Fuel efficiency of the car, measured in miles per gallon (MPG).",
            "properties": {
                "dtype": "number",
                "std": 7.8159843125657815,
                "samples": [
                    19.8,
                    18.0,
                    16.0
                ],
                "nunique": 129,
                "min": 9.0,
                "max": 46.6
            }
        }
    ]
}
\end{lstlisting}

\subsubsection{Insight Generation}
The following prompt is the insight generation part of task execution. Based on the python codes in visualization generation,the agent generate the structured insight with examples.
\begin{Prompt}[Insight Generation]   
   Please provide insight for this task: ``Evaluate the relationship between vehicle weight and fuel efficiency." with the initial visualization and analysis codes. Based on previous data transformation process, you need to conduct insight analysis further to report an important insight. Before generating natural language insight text, you need to specify the column names (data variables) and data\_values in the data involved in the insight, as well as the insight type. Improve insight text based on these. Note that if data variables are used in insight text, use the original column names directly.\\
   Here is an example for your output:
    \begin{lstlisting}[language=json]
{
"text": "There is a correlation between A and B, indicating xx generally exhibit lower B for xx in C.",
"type": "Correlation",
"parameters": "negative",
"data_variables": [`A', `B'],
"data_values": {
`A': `2000 to 5000 lbs',
`B': `10 to 35 MPG',
`C': `xx'
}
}
    \end{lstlisting} 
\end{Prompt}

An output example of insight generation.
\begin{lstlisting}[language=json]
{
"text": "There is a strong negative correlation between Weight_in_lbs and Miles_per_Gallon, indicating heavier cars generally exhibit lower Miles_per_Gallon.",
"type": "Correlation",
"parameters": "negative",
"data_variables": [
    "Weight_in_lbs",
    "Miles_per_Gallon"
],
"data_values": {
    "Weight_in_lbs": "1613 to 5140 lbs",
    "Miles_per_Gallon": "9 to 46.6 MPG"
}
}
\end{lstlisting} 

\subsubsection{Merge Visualization}
For the different selected views, given generates altair and vega-lite codes for these views, merges these views and adds interactive functions between views according to the given interaction rules.
\begin{Prompt}[Merge visualization]

There are some Altair codes:\\
(Add visualization codes)\\
for creating visualizations, please combine them into one chart and add interactive features. Do not change the number of charts in the codes list. Just use the variable 'data' to refer to this dataframe and do not read another data. You should continue use altair to generate linked visualization using given charts. I want these charts can be brushed to filter each other. Please give me the whole codes including data analysis and chart part in a function `plot(data)'. You only need to give the function definition of plot(data), not execute it. please write the codes in this structure:\textasciigrave\textasciigrave\textasciigrave python \textasciigrave\textasciigrave\textasciigrave"\\
\end{Prompt}

An example of generated multiple linked-view.
\begin{lstlisting}[language=python]
import altair as alt
import pandas as pd

def plot(data: pd.DataFrame):
    # Data Analysis part for the first chart
    data['issue_date'] = pd.to_datetime(data['issue_date'])
    # Calculate total quantity of trading activity per day
    trading_activity = data.groupby('issue_date').agg(total_quantity=('quantity', 'sum')).reset_index()
    # Calculate percentage change in total quantity from the previous day
    trading_activity['quantity_change'] = trading_activity['total_quantity'].pct_change()
    
    # Data Analysis part for the second chart
    # Group by issue_date and calculate descriptive statistics for 'amount'
    stats = data.groupby('issue_date')['amount'].describe().reset_index()
    stats = stats.melt(id_vars=['issue_date'], value_vars=['min', '25%', '50%', '75%', 'max'],
                       var_name='statistic', value_name='amount')
    
    # Create a shared selection for interactivity
    shared_brush = alt.selection_interval(encodings=['x'])
    
    # Create the first chart: Trading Activity Over Time
    chart1 = alt.Chart(trading_activity).mark_line().encode(
        x=alt.X('issue_date:T', axis=alt.Axis(title='Issue Date')),
        y=alt.Y('total_quantity:Q', axis=alt.Axis(title='Total Quantity')),
        tooltip=['issue_date:T', 'total_quantity:Q']
    ).properties(
        title='Trading Activity Over Time',
        width=600,
        height=400
    ).add_selection(
        shared_brush
    )
    
    # Create the second chart: Distribution of Transaction Amounts
    bar_chart = alt.Chart(stats).mark_bar().encode(
        x=alt.X('statistic:N', axis=alt.Axis(title='Statistic')),
        y=alt.Y('amount:Q', axis=alt.Axis(title='Amount')),
        color=alt.Color('issue_date:T', legend=None),
        tooltip=['issue_date:T', 'statistic', 'amount']
    ).properties(
        title='Distribution of Transaction Amounts',
        width=200,
        height=400
    ).transform_filter(
        shared_brush
    )
    
    # Combine the charts using hconcat
    combined_chart = alt.hconcat(chart1, bar_chart).resolve_scale(color='shared')
    
    return combined_chart
\end{lstlisting}

\subsubsection{Is Completed}
Based on the task content and the insights generated in the previous step, the agent scores the task according to the degree of completion and gives explanations.
\begin{Prompt}[Is completed]
    For the given task: ``Evaluate the relationship between vehicle weight and fuel efficiency.", and some derived insights: ``There is a strong negative correlation between Weight\_in\_lbs and Miles\_per\_Gallon, indicating heavier cars generally exhibit lower Miles\_per\_Gallon." and codes to solve this task.\\
    You need to judge whether the task requires further analysis. The complexity of the task can be considered from data, data mining, and visualization methods. If the task appears incomplete, it needs to be decomposed further. Please rate the task from 1-10. For example, if the initial solution adequately segments the data or if the task requires advanced statistical analysis. You should output a score and explanation of no more than 20 words for your evaluation.
    You should write the result beginning with \textasciigrave\textasciigrave\textasciigrave json and ending with \textasciigrave\textasciigrave\textasciigrave, and the json data includes a score and an explanation. The output structure is like:
    \begin{lstlisting}[language=json]
{
"score": "",
"explanation": ""
}    
    \end{lstlisting}
\end{Prompt}
The output is:
\begin{lstlisting}[language=json]
{
"score": 7/10,
"explanation": "Metadata extracted and relationship identified, but further segmentation and in-depth analysis could reveal more insights."
}    
\end{lstlisting}

\subsubsection{Whether Decompose}

The agent determines whether the task needs to be decomposed based on the task's completion score and explanation. If so, it returns the decomposition plan, otherwise it returns empty.
\begin{Prompt}[Decompose Planner]
    For the given goal: ``My goal is to analyze the fuel efficiency," and task: ``Evaluate the relationship between vehicle weight and fuel efficiency.", the generated derived some insights: ``There is a strong negative correlation between Weight\_in\_lbs and Miles\_per\_Gallon, indicating heavier cars generally exhibit lower Miles\_per\_Gallon." and visualizations, discriminating whether this task requires further analysis, such as using more detailed data segmentation or advanced statistical methods.\\
    The completion score of this task is 7/10, with the reasons: ``Metadata extracted and relationship identified, but further segmentation and in-depth analysis could reveal more insights.". 
    If needed, please generate no more than 3 subtasks and indicate the methods they use respectively based on this.  You should write the tasks begin with \textasciigrave\textasciigrave\textasciigrave json and end with \textasciigrave\textasciigrave\textasciigrave, and the json data includes a list of tasks like this:
    \begin{lstlisting}[language=json]
{
    "tasks": ["","",""],
    "execution_order": "(Task_1 AND Task_2 AND Task_3)",
    "methods": ["","",""],
    "data variables":[["",""],[],[]],
    "task_type":["","",""]
}    
    \end{lstlisting}
    For data\_variables, the task may not correspond to the data, so you need to go to the data set and find the corresponding data variables, and even if transformation is required, you need to write the original column. There are two operators ``AND" and ``DOWN" in the execution order to connect tasks. ``AND" indicates that all subtasks need to complete successfully but there is no requirement for execution order, ``DOWN" indicates that subtasks need to be executed step by step in order. Each task is a sentence no more than 20 words, methods include data segmentation, statistical methods. If do not need to decompose, output \{``tasks": `null'\}.  No need to say Here is the requested JSON format listing the tasks:.
\end{Prompt}    

An output example of whether decompose.
\begin{lstlisting}[language=json]
{
"tasks": ["Segment data by Origin to compare regional differences in weight and fuel efficiency correlations.","Apply linear regression to model the relationship between weight and fuel efficiency.","Investigate impact of Cylinders on fuel efficiency within different weight categories."],
"execution_order": "(Task_1 AND Task_2 AND Task_3)",
"methods": ["Data segmentation","Regression analysis","Segmentation analysis"],
"data_variables":[["Weight_in_lbs", "Miles_per_Gallon", "Origin"],["Weight_in_lbs","Miles_per_Gallon"],["Cylinders", "Miles_per_Gallon", "Weight_Category"]],
"task_type":["Analysis","Modeling","Investigation"]
}    
\end{lstlisting}
Another two examples of whether decompose. In both examples, the `tasks' represent three subtasks decomposed from the previous task, and the execution order is `AND' and `DOWN', which means that the decomposition tasks are executed in the following logic and order: the first subtask is executed in parallel with the second subtask, i.e., the decomposition task is triggered again when either subtask fails, and then the third subtask is executed in serial order.
\begin{lstlisting}[language=json]
{
"tasks": ["Segment data by Origin to compare regional differences in weight and fuel efficiency.", "Apply linear regression to model the relationship between weight and fuel efficiency.", "Investigate impact of Cylinders on fuel efficiency within different weight categories."],
"execution_order": "(Task_1 AND Task_2 DOWN Task_3)"
}
\end{lstlisting}
\begin{lstlisting}[language=json]
{
"tasks": ["Investigate polynomial relationships between weight and MPG for each cylinder category.", "Model interaction terms between cylinders, horsepower, and displacement to assess combined effects on MPG.", "Perform residual analysis to evaluate model fit and discover potential outlier influence."],
"execution_order": "(Task_1 AND Task_2 DOWN Task_3)"
}
\end{lstlisting}

\subsubsection{Merge Visualization}
For the different selected views, the given generates Altair and Vega-Lite codes for these views, merges these views, and adds interactive functions between views according to the given interaction rules.
\begin{Prompt}[Merge visualization]

There are some Altair codes:\\
(Add visualization codes)\\
for creating visualizations, please combine them into one chart and add interactive features. Do not change the number of charts in the codes list. Just use the variable `data' to refer to this dataframe and do not read other data. You should continue to use Altair to generate linked visualization using the given charts. I want these charts can be brushed to filter each other. 
First, you should ensure both charts contain common key columns with consistent data formats for the selection fields.
Second, you should add selections based on the chart type, such as a brush on the time axis for the line chart, dual brushes for the scatter plot, and click interactions for the bar chart.
Third, the layout should be no more than three charts in a row.
Please give me the whole code, including the data analysis and chart part, in a function `plot(data)'. You only need to give the function definition of the plot(data), not execute it. please write the codes in this structure:\textasciigrave\textasciigrave\textasciigrave python \textasciigrave\textasciigrave\textasciigrave"\\
\end{Prompt}

\subsubsection{Summarize Insights}
Given a batch of the same task, decompose it into subtasks and their insights and execution orders, and summarize their insights.
\begin{Prompt}[Summarize Insights]
    Now the decomposed sub-tasks for the task ``Evaluate the relationship between vehicle weight and fuel efficiency." are executed. The decomposition logic is ``(Task\_1 AND Task\_2 AND Task\_3)". You need to give a summary insight for the source task based on insight from each subtask. Here is the results for each sub-task: ``There is a strong negative correlation between Weight\_in\_lbs and Miles\_per\_Gallon, indicating heavier cars generally exhibit lower Miles\_per\_Gallon.". Your output should be begin with \textasciigrave\textasciigrave\textasciigrave json and end with \textasciigrave\textasciigrave\textasciigrave, like this:
    \begin{lstlisting}[language=json]
{"summary_insight":""}.
    \end{lstlisting}
\end{Prompt}

The output example of merge visualization.
\begin{lstlisting}[language=json]
{"summary_insight":"Vehicle weight negatively impacts fuel efficiency, with the greatest effects seen in heavier, USA-made cars. A linear model quantified this decrease as -0.0077 MPG per pound. Additionally, increasing engine cylinders reduces MPG across all weight categories, more so in heavier vehicles."}.
\end{lstlisting}

\subsection{Task Flow}
\label{appsec:appendix_history}
Analysis history stores the path of user historical exploration. It is a task-driven directed graph flow. Each node in the graph represents a task and the edges represent the relationship between tasks. There are three types of nodes: goal, task, subtask, and the edges include Decomposition, enlightened, and review.
Subtask refers to the node obtained by task decomposition. The edge type connected from the parent task node to the child subtask node is decomposition. The edge of recommendation represents the agent recommending new tasks, and the edge of review represents the agent recommending review of nodes that have been proposed but not explored.
We provide an analysis history example.
\begin{lstlisting}[language=json]
{
  "nodes": [
    {
        "id": 1,
        "group": 1,
        "task": "My goal is to analyze influencing factors of fuel efficiency\n",
        "type": "goal",
        "variables": null,
        "task_type": null,
        "visualization": null,
        "insight": null,
        "refined_insight": null,
        "summary_insight": null,
        "chat_text": "The dataset contains 406 entries of vehicle data, including name, MPG, engine attributes, weight, acceleration, year, and origin. It has some missing values and requires date-time conversion for the 'Year' column.",
        "error_message": null,
        "execution_order_list": null,
        "progress": 0.5,
        "codes": null
    },
    {
        "id": 2,
        "group": 1,
        "task": "Analyze the distribution of fuel efficiency across the dataset.",
        "type": "task",
        "variables": [
            "Miles_per_Gallon"
        ],
        "task_type": "distribution",
        "visualization": null,
        "insight": null,
        "refined_insight": null,
        "summary_insight": null,
        "chat_text": null,
        "error_message": null,
        "execution_order_list": null,
        "progress": 0.0,
        "codes": null
    },
    {
        "id": 3,
        "group": 0,
        "task": "Evaluate the relationship between vehicle weight and fuel efficiency.",
        "type": "task",
        "variables": [
            "Weight_in_lbs",
            "Miles_per_Gallon"
        ],
        "task_type": "correlation",
        "visualization": {
          "$schema": "https://vega.github.io/schema/vega-lite/v5.15.1.json",
          "config": {
              "view": {
                  "continuousHeight": 300,
                  "continuousWidth": 300
              }
          },
          "data": {
              "url": "http://127.0.0.1:5000/data/data_507d1eff-14b2-426d-b87a-b8cf18c9c741.json"
          },
          "height": 400,
          "layer": [
              {
                  "encoding": {
                      "color": {
                          "field": "Origin",
                          "legend": {
                              "title": "Origin"
                          },
                          "type": "nominal"
                      },
                      "opacity": {
                          "condition": {
                              "param": "param_3",
                              "value": 0.9
                          },
                          "value": 0.2
                      },
                      "tooltip": [
                          {
                              "field": "Name",
                              "type": "nominal"
                          },
                          {
                              "field": "Weight_in_lbs",
                              "type": "quantitative"
                          },
                          {
                              "field": "Miles_per_Gallon",
                              "type": "quantitative"
                          },
                          {
                              "field": "Year",
                              "type": "temporal"
                          },
                          {
                              "field": "Origin",
                              "type": "nominal"
                          }
                      ],
                      "x": {
                          "axis": {
                              "title": "Vehicle Weight (lbs)"
                          },
                          "field": "Weight_in_lbs",
                          "type": "quantitative"
                      },
                      "y": {
                          "axis": {
                              "title": "Fuel Efficiency (MPG)"
                          },
                          "field": "Miles_per_Gallon",
                          "type": "quantitative"
                      }
                  },
                  "mark": {
                      "size": 60,
                      "type": "circle"
                  },
                  "name": "view_3"
              },
              {
                  "encoding": {
                      "color": {
                          "field": "Origin",
                          "legend": {
                              "title": "Origin"
                          },
                          "type": "nominal"
                      },
                      "opacity": {
                          "condition": {
                              "param": "param_3",
                              "value": 0.9
                          },
                          "value": 0.2
                      },
                      "size": {
                          "condition": {
                              "test": {
                                  "not": {
                                      "param": "param_3"
                                  }
                              },
                              "value": 30
                          },
                          "value": 60
                      },
                      "tooltip": [
                          {
                              "field": "Name",
                              "type": "nominal"
                          },
                          {
                              "field": "Weight_in_lbs",
                              "type": "quantitative"
                          },
                          {
                              "field": "Miles_per_Gallon",
                              "type": "quantitative"
                          },
                          {
                              "field": "Year",
                              "type": "temporal"
                          },
                          {
                              "field": "Origin",
                              "type": "nominal"
                          }
                      ],
                      "x": {
                          "axis": {
                              "title": "Vehicle Weight (lbs)"
                          },
                          "field": "Weight_in_lbs",
                          "type": "quantitative"
                      },
                      "y": {
                          "axis": {
                              "title": "Fuel Efficiency (MPG)"
                          },
                          "field": "Miles_per_Gallon",
                          "type": "quantitative"
                      }
                  },
                  "mark": {
                      "size": 60,
                      "type": "circle"
                  }
              }
          ],
          "params": [
              {
                  "name": "param_3",
                  "select": {
                      "encodings": [
                          "x"
                      ],
                      "type": "interval"
                  },
                  "views": [
                      "view_3"
                  ]
              },
              {
                  "name": "param_4",
                  "select": {
                      "fields": [
                          "Name"
                      ],
                      "type": "point"
                  },
                  "views": [
                      "view_3"
                  ]
              }
          ],
          "resolve": {
              "scale": {
                  "color": "independent",
                  "size": "independent"
              }
          },
          "width": 600
      },
        "insight": {
            "text": "There is a strong negative correlation between Weight_in_lbs and Miles_per_Gallon, indicating heavier cars generally exhibit lower Miles_per_Gallon.",
            "type": "Correlation",
            "parameters": "negative",
            "data_variables": [
                "Weight_in_lbs",
                "Miles_per_Gallon"
            ],
            "data_values": {
                "Weight_in_lbs": "1613 to 5140 lbs",
                "Miles_per_Gallon": "9 to 46.6 MPG"
            }
        },
        "refined_insight": null,
        "summary_insight": "Vehicle weight negatively impacts fuel efficiency, with the greatest effects seen in heavier, USA-made cars. A linear model quantified this decrease as -0.0077 MPG per pound. Additionally, increasing engine cylinders reduces MPG across all weight categories, more so in heavier vehicles.",
        "chat_text": null,
        "error_message": "null",
        "execution_order_list": [
            "AND",
            "DOWN"
        ],
        "progress": 1,
        "codes": "import altair as alt\nimport pandas as pd\n\ndef plot(data: pd.DataFrame):\n    # Ensure correct data types\n    data = data.astype({\n        'Weight_in_lbs': float,\n        'Miles_per_Gallon': float\n    }).dropna(subset=['Weight_in_lbs', 'Miles_per_Gallon'])  # Remove rows with missing values\n\n    # Create selection features\n    click = alt.selection_multi(fields=['Name'])\n    brush = alt.selection(type='interval', encodings=['x'])\n\n    # Base chart for scatter plot\n    scatter = alt.Chart(data).mark_circle(size=60).encode(\n        x=alt.X('Weight_in_lbs', axis=alt.Axis(title='Vehicle Weight in lbs')),\n        y=alt.Y('Miles_per_Gallon', axis=alt.Axis(title='Fuel Efficiency (MPG)')),\n        color=alt.condition(brush, 'Origin:N', alt.value('lightgray')),\n        opacity=alt.condition(click, alt.value(0.9), alt.value(0.2)),\n        tooltip=['Name:N', 'Weight_in_lbs', 'Miles_per_Gallon', 'Year:T', 'Origin:N']\n    ).add_selection(\n        brush\n    ).add_selection(\n        click\n    ).properties(\n        width=600,\n        height=400\n    )\n\n    # Base chart for histogram\n    histogram = alt.Chart(data).mark_bar().encode(\n        x=alt.X('count()', axis=alt.Axis(title='Number of Cars')),\n        y=alt.Y('Origin:N', axis=alt.Axis(title='Country of Origin')),\n        color=alt.Color('Origin:N', legend=alt.Legend(title=\"Origin\")),\n        opacity=alt.condition(click, alt.value(0.9), alt.value(0.2)),\n    ).transform_filter(\n        brush\n    ).add_selection(\n        click\n    ).properties(\n        width=600\n    )\n\n    return alt.vconcat(scatter, histogram)"
    },
    {
        "id": 7,
        "group": 0,
        "task": "Segment data by Origin to compare regional differences in weight and fuel efficiency correlations.",
        "type": "sub_task",
        "variables": [
            "Weight_in_lbs",
            "Miles_per_Gallon",
            "Origin"
        ],
        "task_type": "Analysis",
        "visualization": null,
        "insight": null,
        "refined_insight": null,
        "summary_insight": null,
        "chat_text": null,
        "error_message": null,
        "execution_order_list": null,
        "progress": 0.0,
        "codes": null
    },
    {
        "id": 8,
        "group": 0,
        "task": "Apply linear regression to model the relationship between weight and fuel efficiency.",
        "type": "sub_task",
        "variables": [
            "Weight_in_lbs",
            "Miles_per_Gallon"
        ],
        "task_type": "Modeling",
        "visualization": null,
        "insight": null,
        "refined_insight": null,
        "summary_insight": null,
        "chat_text": null,
        "error_message": null,
        "execution_order_list": null,
        "progress": 0.0,
        "codes": null
    }
    
],
"edges":[
    {
        "source": 1,
        "target": 2,
        "weight": 1.0,
        "type": "Enlightenment",
        "labels": [],
        "differences": {
            "changes": {
                "variables_change": {
                    "original": [],
                    "new": [
                        "Miles_per_Gallon"
                    ]
                },
                "type_change": {
                    "original": [
                        "goal"
                    ],
                    "new": [
                        "task"
                    ]
                },
                "insight_change": {
                    "original": [],
                    "new": [
                        "DISTRIBUTION"
                    ]
                }
            },
            "costs": {
                "variables_cost": 1.0,
                "type_cost": 1.0,
                "insight_cost": 1.0,
                "total_difference_cost": 1.0
            }
        }
    },
    {
        "source": 1,
        "target": 3,
        "weight": 1.0,
        "type": "Enlightenment",
        "labels": [],
        "differences": {
            "changes": {
                "variables_change": {
                    "original": [],
                    "new": [
                        "Weight_in_lbs",
                        "Miles_per_Gallon"
                    ]
                },
                "type_change": {
                    "original": [
                        "goal"
                    ],
                    "new": [
                        "task"
                    ]
                },
                "insight_change": {
                    "original": [],
                    "new": [
                        "CORRELATION",
                        "CORRELATION"
                    ]
                }
            },
            "costs": {
                "variables_cost": 1.0,
                "type_cost": 1.0,
                "insight_cost": 1.0,
                "total_difference_cost": 1.0
            }
        }
    },
    {
        "source": 3,
        "target": 4,
        "weight": 0.2760920571391645,
        "type": "Decomposition",
        "labels": [],
        "differences": {
            "changes": {
                "variables_change": {
                    "original": [
                        "Weight_in_lbs",
                        "Miles_per_Gallon"
                    ],
                    "new": [
                        "Weight_in_lbs",
                        "Miles_per_Gallon",
                        "Origin"
                    ]
                },
                "type_change": {
                    "original": [
                        "task"
                    ],
                    "new": [
                        "sub_task"
                    ]
                },
                "insight_change": {
                    "original": [
                        "CORRELATION",
                        "CORRELATION"
                    ],
                    "new": [
                        "CORRELATION",
                        "ANALYSIS"
                    ]
                }
            },
            "costs": {
                "variables_cost": 0.18350341907227408,
                "type_cost": 1.0,
                "insight_cost": 0.29289321881345254,
                "total_difference_cost": 0.2760920571391645
            }
        }
    },
    {
        "source": 3,
        "target": 5,
        "weight": 0.20000000000000018,
        "type": "Decomposition",
        "labels": [],
        "differences": {
            "changes": {
                "variables_change": {
                    "original": [
                        "Weight_in_lbs",
                        "Miles_per_Gallon"
                    ],
                    "new": [
                        "Weight_in_lbs",
                        "Miles_per_Gallon"
                    ]
                },
                "type_change": {
                    "original": [
                        "task"
                    ],
                    "new": [
                        "sub_task"
                    ]
                },
                "insight_change": {
                    "original": [
                        "CORRELATION",
                        "CORRELATION"
                    ],
                    "new": [
                        "REGRESSION ANALYSIS",
                        "MODELING"
                    ]
                }
            },
            "costs": {
                "variables_cost": 2.220446049250313e-16,
                "type_cost": 1.0,
                "insight_cost": 1.0,
                "total_difference_cost": 0.20000000000000018
            }
        }
    },
    {
        "source": 5,
        "target": 2,
        "weight": 0.0,
        "type": "Review",
        "labels": [],
        "differences": "Node(s) not found"
    }
]
}
\end{lstlisting}

%% file: sections/appendix-test-llm.tex
\section{LLM Performance Test}
\label{appsec:appendix_test}

This appendix analyzes the performance of GPT-3.5-turbo and GPT-4-turbo on data analysis and visualization tasks. We tested 160 tasks across two datasets, focusing on error types and model efficiency. The results highlight common error categories and compare the models' performance in terms of error rates and time costs.

\subsection{Error Analysis}

To enhance the robustness of our system, we conducted tests focusing on 2 datasets, with 20 tasks per dataset, utilizing GPT-3.5-turbo and GPT-4-turbo. 
Each task was tested with data analysis code generation and insight annotation. Both of these two steps tested two prompts to support the choice of better results. 
This setting resulted in a total of 
\textit{ 2 \text{(datasets)} $\times$ 10 \text{(tasks per dataset)} $\times$ 2 \text{(models)} $\times$ 2 \text{(codes generation prompts)} $\times$ 2 \text{(insight annotation prompts)} = 160} initial results.

To improve error handling, we allow the model to self-debug again while we prompt the model with cached error messages, yielding an overall 204 experimental results.
In the first round, out of the 160 tasks, 44 errors were identified. Upon a second verification, 18 of these errors were resolved, while 26 remained erroneous.

\subsubsection{Error Types}

\begin{figure}[ht]
    \centering
    \includegraphics[width=\linewidth]{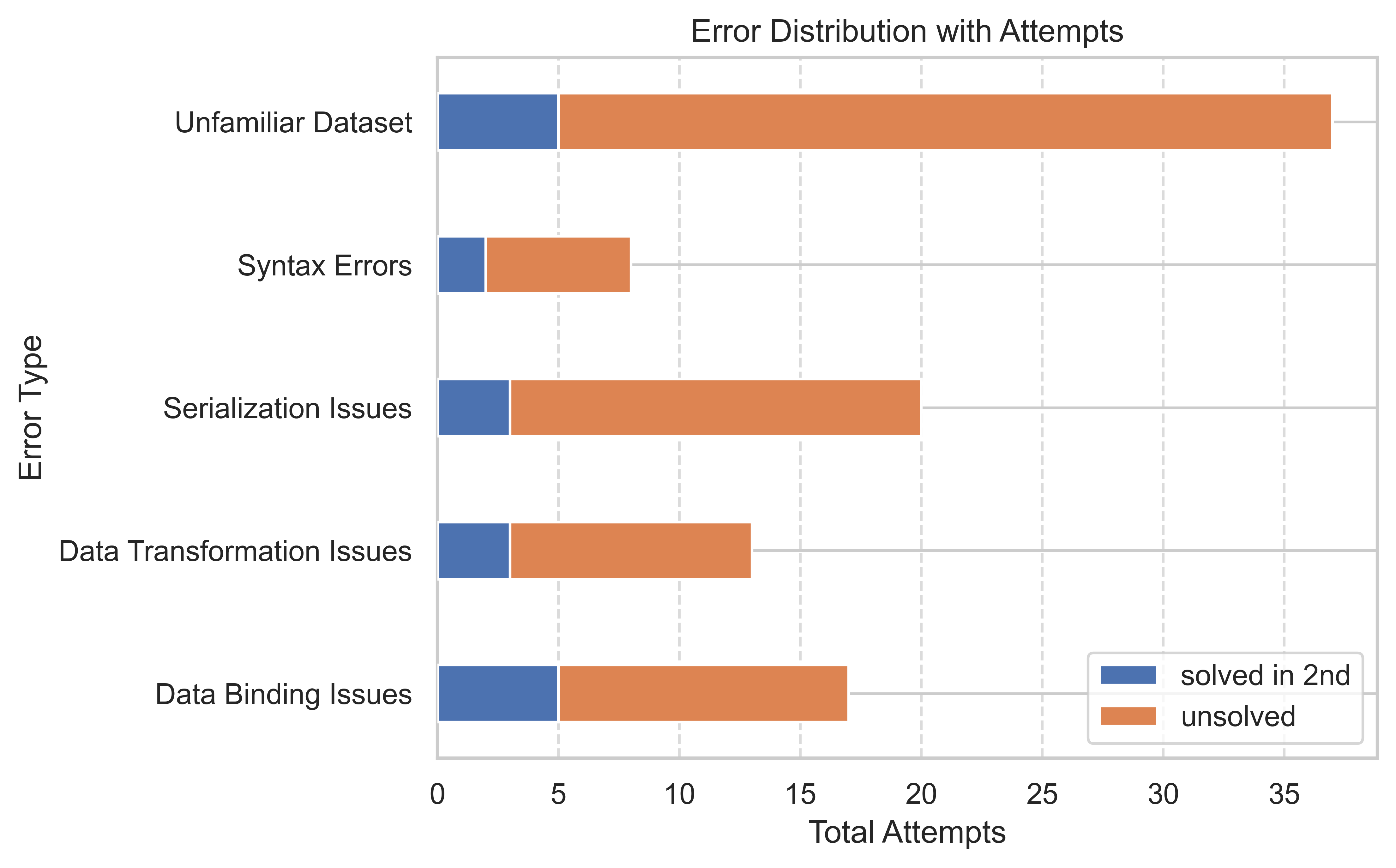}
    \vspace{-0.7cm}
    \caption{Distribution of error types encountered across various attempts, categorized by whether the issue was resolved in a second attempt or remained unsolved. Error types include issues related to unfamiliar datasets, syntax, serialization, data transformation, and data binding.}
    \label{fig:error-type}
\end{figure}

We classified these 70 ($44+26$) errors we found into five categories, as illustrated in Fig.~\ref{fig:error-type}.

\begin{itemize}[leftmargin=*]
    \item \textbf{Unfamiliar Dataset:} These errors highlight the model's lack of understanding regarding the structure and methods applicable to the given dataset. The model attempts to use non-existent attributes or incorrect methods, indicating a gap in its knowledge about the dataset's schema and the operations that can be performed on it.
    \begin{itemize}[leftmargin=*]
        \item \schemaname{`DataFrame' object has no attribute `reset\_src\_index'}
        \item \schemaname{Cannot reset\_index inplace on a Series to create a DataFrame}
    \end{itemize}

    \item \textbf{Data Binding Issues:} These errors occur when the model incorrectly references variable names that do not exist in the given context. This suggests that the model has difficulty maintaining consistency in variable naming and ensuring that all referenced variables are defined and accessible within the code.
    \begin{itemize}[leftmargin=*]
        \item \schemaname{name `actuator\_line' is not defined}
        \item \schemaname{name `hp\_preusline' is not defined}
    \end{itemize}

    \item \textbf{Serialization Issues:} These errors arise when the model attempts to use data types that are not compatible with JSON serialization, which is required for certain operations such as visualization with Altair. These errors indicate that the model does not adequately understand the requirements for data serialization and the need to use JSON-compatible data types.
    \begin{itemize}[leftmargin=*]
        \item \schemaname{Object of type Interval is not JSON serializable}
        \item \schemaname{Object of type Period is not JSON serializable}
    \end{itemize}

    \item \textbf{Data Transformation Issues:} These errors indicate the model's difficulty in converting data from one type to another and identifying correct data types. The first example demonstrates a failure to convert a string of concatenated names into a numeric format, while the second highlights the challenge of correctly identifying and spelling field names.
    \begin{itemize}[leftmargin=*]
    \item \schemaname{Could not convert ... to numeric}
    \item \schemaname{Unable to determine data type for the field "Origin"}
    \end{itemize}

    \item \textbf{Syntax Errors:} Syntax errors occur when the model generates code that does not adhere to the correct syntax rules of the programming language. These errors suggest that the model may overlook essential syntax elements such as matching parentheses or correctly structuring the code.
    \begin{itemize}[leftmargin=*]
        \item \schemaname{invalid syntax (<string>, line 37)}
    \end{itemize}
\end{itemize}

\subsubsection{Comparison of Models}

Based on the analysis of error type distribution with model comparison. we found that:

\textbf{GPT-3.5-turbo has lower overall error counts}: In terms of these code generation task types, the overall \textit{error rate ratio} of GPT-3.5-turbo and GPT-4-turbo is 1:2. This indicates that GPT-4-turbo commits more errors overall compared to GPT-3.5-turbo, as shown by the larger counts of unsolved errors.

\begin{figure}[ht]
    \centering
    \includegraphics[width=\linewidth]{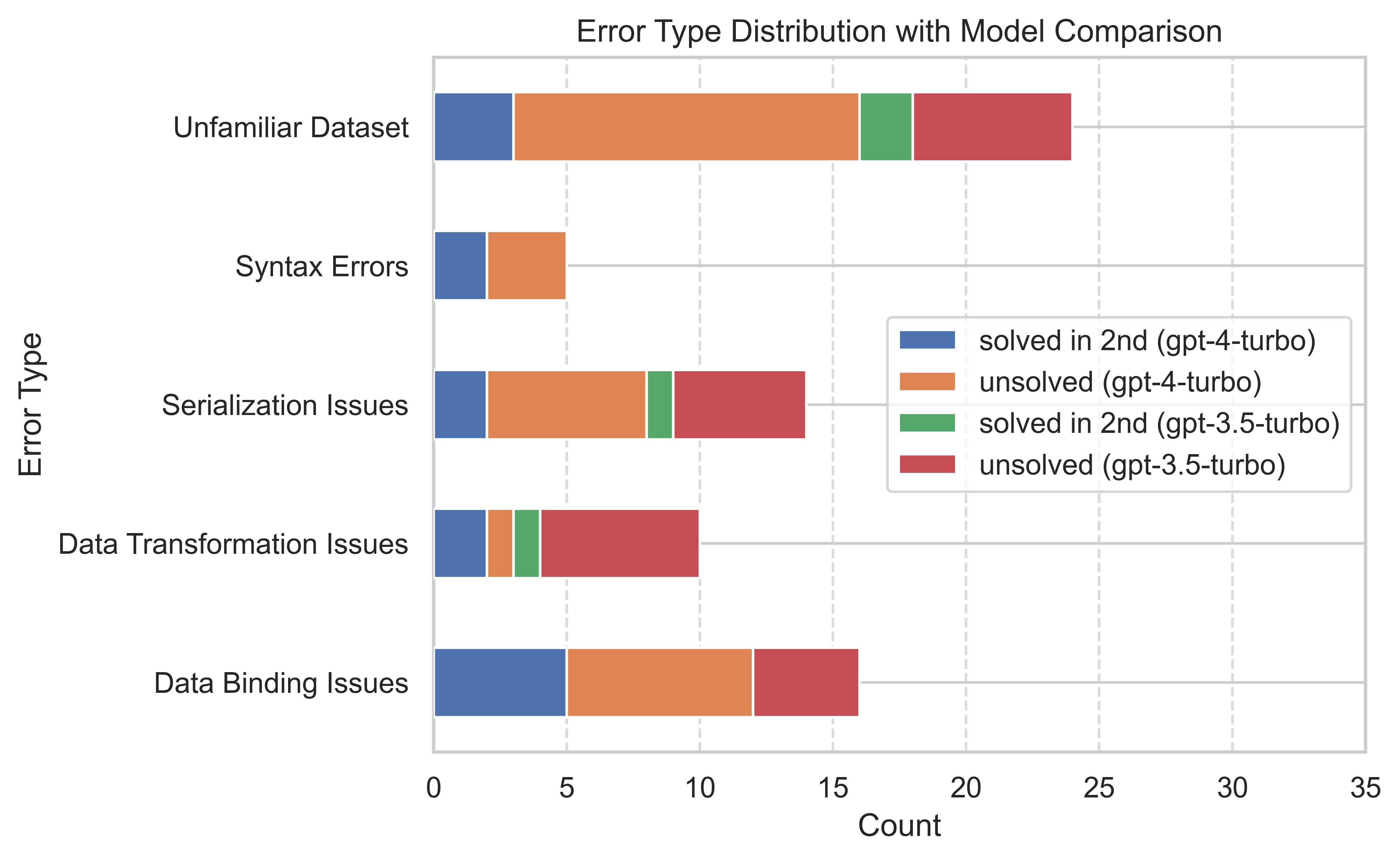}
    \vspace{-0.7cm}
    \caption{Distribution of error types across two models (GPT-4-turbo and GPT-3.5-turbo), showing the count of errors resolved in a second attempt or left unsolved.}
    \label{fig:error-model}
\end{figure}

\textbf{Repair Ability of GPT-4-turbo}: Despite the higher error rate, GPT-4-turbo shows significant repair ability in the second round, solving many errors that were initially unsolved (\Cref{fig:error-model}). This repair ability is evident across all error types, particularly in \textit{Unfamiliar Dataset} and \textit{Data Binding Issues}.

\subsubsection{Comparison of Tasks}

The bar chart illustrates the relationship between various \textit{task types} and specific \textit{error types} encountered in data analysis and visualization using large models. The tasks for the tests are shown in~\Cref{tab:tasks_analysis}.

\begin{table*}[t]
\centering
\captionsetup{font=normal}
\caption{Tasks Analysis: The tasks are derived from the datasets used in the evaluation and proposed by the agent. The \textit{``Cars''} dataset is analyzed in the first ten tasks, while the \textit{``Superstore''} dataset is analyzed in the subsequent ten tasks. The types of tasks are labeled by the LLM.}

\label{tab:tasks_analysis}
\begin{tabularx}{\textwidth}{lXl}
\toprule
dataset & task & type \\
\midrule
Cars & Analyze the distribution of fuel efficiency across the dataset. & Distribution Analysis \\
Cars & Evaluate the relationship between vehicle weight and fuel efficiency. & Correlation Analysis \\
Cars & Compare the average horsepower for vehicles of different origins. & Comparative Analysis \\
Cars & Examine how vehicle characteristics have changed over time. & Trend Analysis \\
Cars & Assess the distribution of engine cylinders in the dataset. & Distribution Analysis \\
Cars & Segment data by Origin to compare regional differences in weight and fuel efficiency correlations. & Segmentation Analysis \\
Cars & Apply linear regression to model the relationship between weight and fuel efficiency. & Predictive Modeling \\
Cars & Investigate the impact of Cylinders on fuel efficiency within different weight categories. & Correlation Analysis \\
Cars & Analyze the variance in MPG within each cylinder count across weight categories. & Variance Analysis \\
Cars & Perform a cluster analysis to identify groups of cars with similar weight and fuel efficiency profiles. & Cluster Analysis \\
Superstore & Analyze the correlation between sales and profit to understand how they are related at a product or category level. & Correlation Analysis \\
Superstore & Aggregate sales by category and sub-category to identify which product groups are performing best. & Aggregation Analysis \\
Superstore & Examine the distribution of sales and profit to determine common ranges and any outliers or anomalies in the data. & Distribution and Outlier Analysis \\
Superstore & Compare average discount rates across different segments (Consumer, Corporate, Home Office) to assess the impact of discounts on sales and profit. & Correlation Analysis \\
Superstore & Segment monthly sales data by product categories. & Segmentation Analysis \\
Superstore & Analyze sales trends of top-selling and bottom-selling products. & Trend Analysis \\
Superstore & Perform cluster analysis to group similar products based on sales trends and profitability. & Cluster Analysis \\
Superstore & Investigate the impact of promotions or discounts on monthly sales trends by product category. & Correlation Analysis \\
Superstore & Determine the product groups with the highest profit margin within each category. & Margin Analysis \\
Superstore & Analyze the impact of seasonal trends on profit margins by category and sub-category. & Trend Analysis \\
\bottomrule
\end{tabularx}
\end{table*}

For \textit{Correlation Analysis}, the primary errors are \textit{Data Transformation Issues} and \textit{Data Binding Issues}, with occasional \textit{Syntax Errors}. In \textit{Trend Analysis}, the most frequent issues are \textit{Data Transformation Issues}, indicating challenges in converting data formats or structures, and \textit{Data Binding Issues}, which point to problems in linking data with visualization elements.

\textit{Segmentation Analysis} also suffers from \textit{Data Transformation Issues} and \textit{Data Binding Issues}. For \textit{Predictive Modeling}, significant errors include \textit{Serialization Issues}, suggesting difficulties in data serialization processes, along with \textit{Data Binding Issues}.

\textit{Variance Analysis} encounters substantial \textit{Unfamiliar Dataset} issues, highlighting challenges in working with new or unexpected data, and occasional \textit{Serialization Issues}. \textit{Cluster Analysis} is predominantly affected by \textit{Unfamiliar Dataset} issues, reflecting problems with novel dataset, and some \textit{Data Transformation Issues}.

\textit{Aggregation Analysis} shows minimal errors with an even distribution across different types, while \textit{Distribution Analysis} is mainly affected by \textit{Data Binding Issues}. \textit{Margin Analysis} exhibits few errors, with \textit{Unfamiliar Dataset} and \textit{Data Binding Issues} being the most frequent.

Overall, \textit{Unfamiliar Dataset} and \textit{Data Binding Issues} are the most prevalent errors across various task types, underscoring the challenges in handling novel data and effectively linking data with visualization components.

\begin{figure}[ht]
    \centering
    \includegraphics[width=\linewidth]{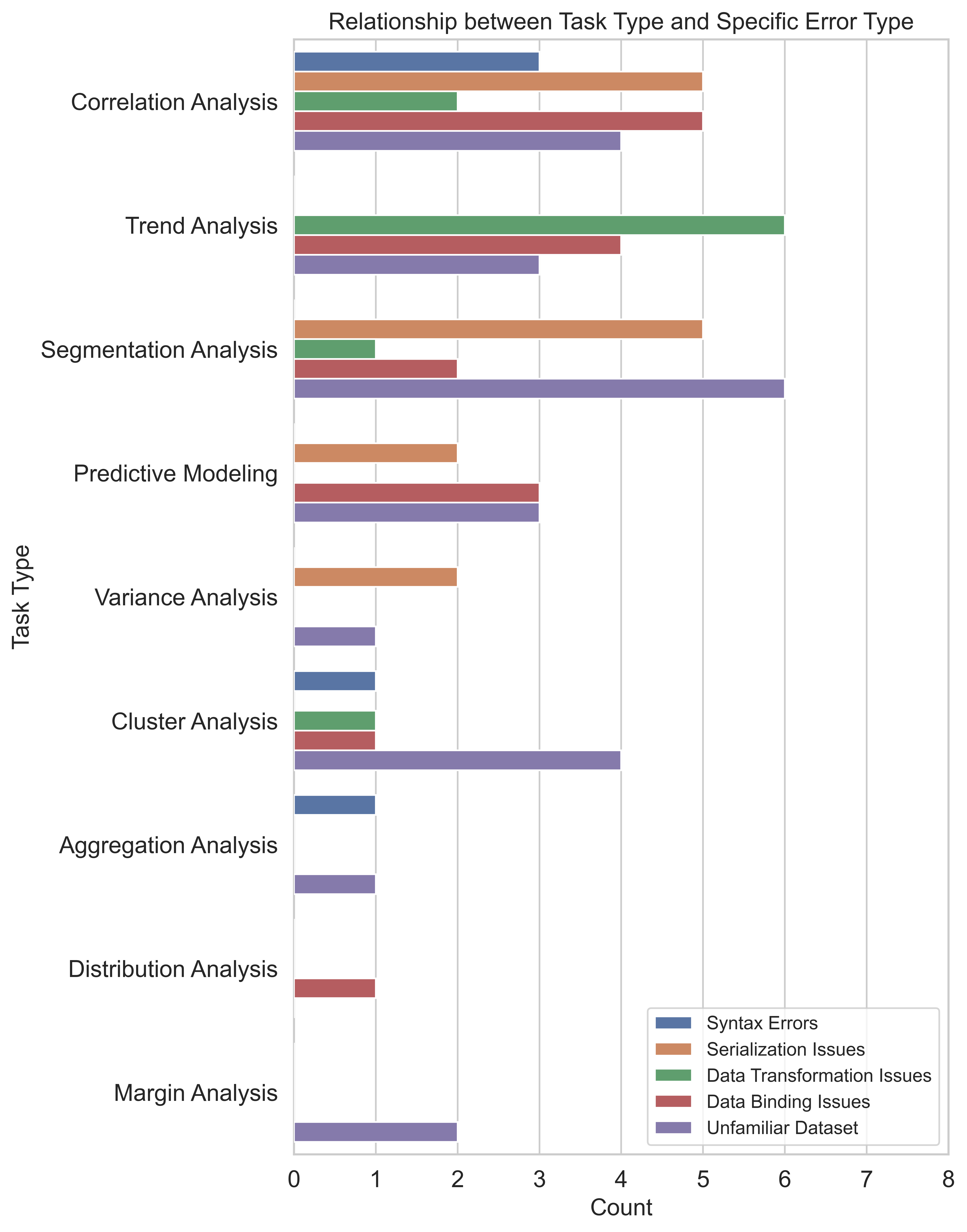}
    \vspace{-0.7cm}
    \caption{Distribution of error types across various analytical tasks, highlighting the susceptibility of each task type (e.g., correlation, trend, segmentation) to specific errors like syntax issues, data binding problems, and unfamiliar datasets.}
    \label{fig:task_error}
    \vspace{-10pt}
\end{figure}

Here's two generated codes and errors for \textit{Trend Analysis} and For \textit{Correlation Analysis}:

\begin{itemize}[leftmargin=*]
    \item \textit{Trend Analysis} has the highest occurrence of \textit{Data Binding Issues}, suggesting frequent difficulties in linking data attributes correctly for trend-related tasks.

    \begin{itemize}[leftmargin=*]
        \item Line 21: \schemaname{name 'top\_slowing' is not defined}
    \end{itemize}
    
    \begin{lstlisting}[language=python]
import altair as alt
import pandas as pd

def plot(data: pd.DataFrame):
# Data Analysis: Calculate Total Sales and Total Profit for each Product
product_summary = data.groupby('Product Name').agg(
    Total_Sales=pd.NamedAgg(column='Sales', aggfunc='sum'),
    Total_Profit=pd.NamedAgg(column='Profit', aggfunc='sum'),
    Total_Quantity=pd.NamedAgg(column='Quantity', aggfunc='sum')
).reset_index()

# Sort by Total Sales to find Top Selling and Bottom Selling Products
top_selling = product_summary.sort_values(by='Total_Sales', ascending=False).head()
bottom_selling = product_summary.sort_values(by='Total_Sales', ascending=True).head()

# Adding annotations: profit for top and bottom selling products
top_selling['Annotation'] = 'Top Selling'
bottom_selling['Annotation'] = 'Bottom Selling'

# Combine the data for top and bottom selling for visualization
annotated_data = pd.concat([top_slowing, bottom_selling], axis=0)

# Chart part
brush = alt.selection_interval(encodings=['x', 'y'], name='Brush select')
chart = alt.Chart(annotated_data).mark_bar().encode(
    x=alt.X('Product Name:N', sort=None, title='Product Name'),
    y=alt.Y('Total_Sales:Q', title='Total Sales'),
    color=alt.Color(`Annotation:N', legend=alt.Legend(title=``Product Category")),
    tooltip=[`Product Name', `Total_Sales', `Total_Profit', `Total_Quantity']
).add_selection(
    brush
).transform_filter(
    brush
)
return chart
        \end{lstlisting}

    \item \textit{Correlation Analysis} frequently encounters \textit{Syntax Errors} and \textit{Data Transformation Issues}, indicating common problems with code syntax and data type handling in these tasks.
    
    \begin{itemize}[leftmargin=*]
        \item \schemaname{invalid syntax (<string>, line 37)}
    \end{itemize}
    \begin{lstlisting}[language=python]
import altair as alt
import pandas as pd

def plot(data: pd.DataFrame):

    # Data Analysis part
    # Perform linear regression using Altair's transform_regression
    regression = alt.transform_regression(
        `Weight_in_lbs', `Miles_per_Gallon', method=``linear"
    )
    
    # Chart part
    brush = alt.selection_interval(encodings=['x', 'y'], name='brush')
    
    # Points for the scatter plot - brushable
    points = alt.Chart(data).mark_point(color='blue').encode(
        x='Weight_in_lbs:Q',
        y='Miles_per_Gallon:Q',
        tooltip=['Name', 'Weight_in_lbs', 'Miles_per_Gallon']
    ).add_selection(
        brush
    ).transform_filter(
        brush
    )
    
    # Line for linear regression - highlighted by brush
    trendline = alt.Chart(data).transform_regression(
        'Weight_in_lbs', 'Miles_per_Gallon', method='linear'
    ).mark_line(color='red').encode(
        x='Weight_in_lbs:Q',
        y='Miles_per_Gallon:Q'
    ).transform_filter(
        brush
    )
    
    # Combining both the points and trend line
    chart = points + trend aoxnli
    
    return chart
        \end{lstlisting}
        
\end{itemize}

\subsection{Visualization Annotations}

To further assess the model's ability to generate visualizations, such as highlighting insights on visualizations, which is more difficult than drawing an overview, we tested these two different situations. 
By comparing the \textit{error rate} with or without annotation requirements, we find that \textit{adding annotations} does increase the probability of errors, especially making errors from unfamiliar datasets and data binding.

\begin{figure}[ht]
    \centering
    \includegraphics[width=\linewidth]{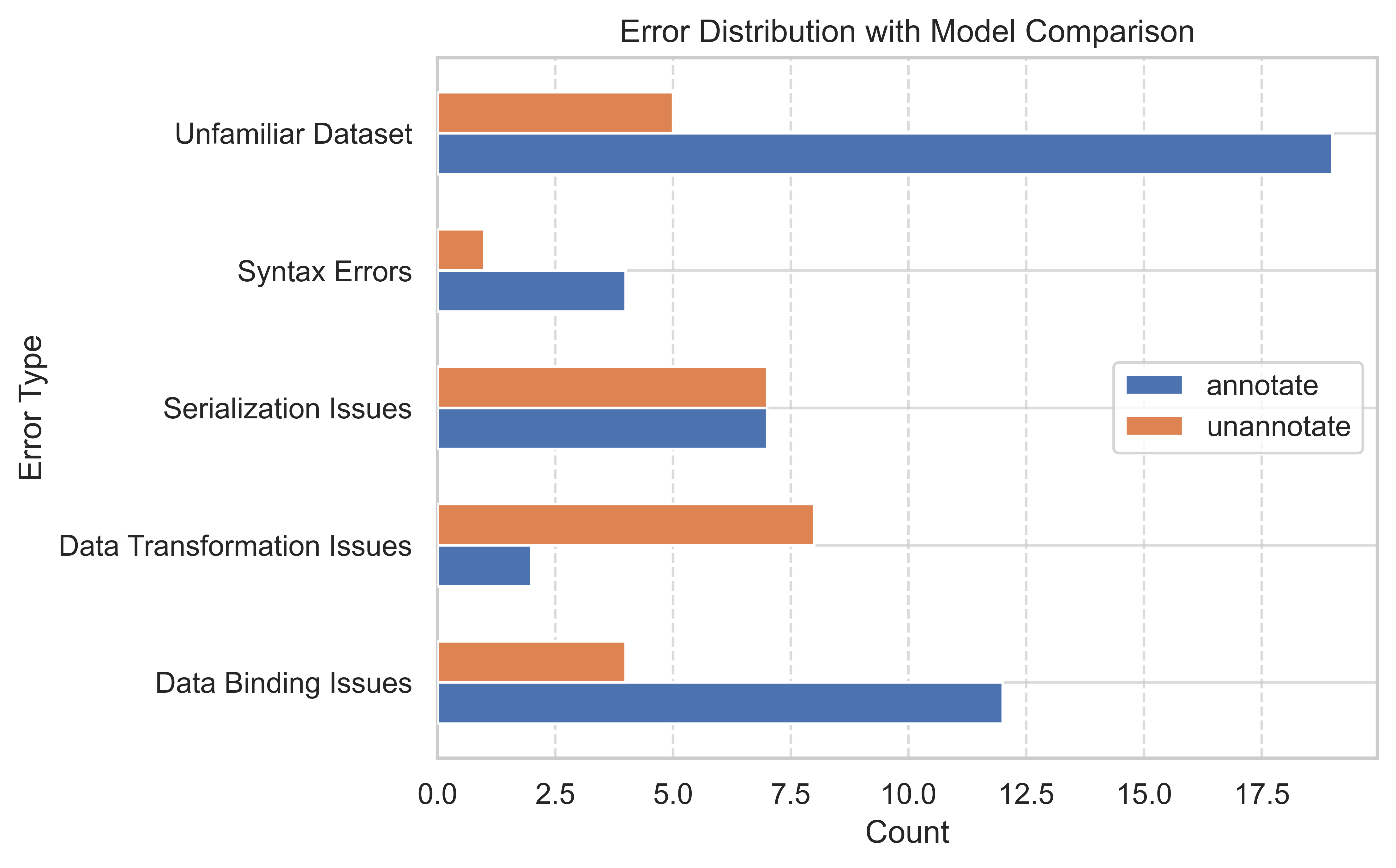}
    \vspace{-0.7cm}
    \caption{Error rates analysis by visualization prompts (annotated or non-annotated). }
    \label{fig:annotate_error}
\end{figure}

Specifically, for data binding issues, there are 12 errors with annotations compared to 4 without, indicating that the complexity introduced by annotations increases the likelihood of errors. Similarly, for unfamiliar datasets, the model makes 19 errors with annotations but only 5 without, suggesting that annotations make it harder for the model to handle unfamiliar data. On the other hand, for data transformation issues, the model performs better with annotations, making only 2 errors compared to 8 without. Serialization issues show an equal distribution of 7 errors for both annotated and unannotated cases, while syntax errors are slightly higher with annotations (4 errors) compared to without (1 error). Results highlight the model's difficulty in managing additional annotation complexity.

\subsection{Insight Generation}
Among all 160 tasks, after self-debugging, 134 successfully generated insight text content that provided some level of analysis of the tasks. However, among these, 18 did not generate the insight in the required JSON format.

We conducted a statistical analysis to compare the success rates of generating each key-value pair in the standardized insight dictionaries.

To generate the insights, we employed two different approaches. The first approach involved generating the insight in JSON format first, while the second approach generated the natural language text first. The following are the results of these two different methods.
\begin{figure}[ht]
    \centering
    \includegraphics[width=0.8\linewidth]{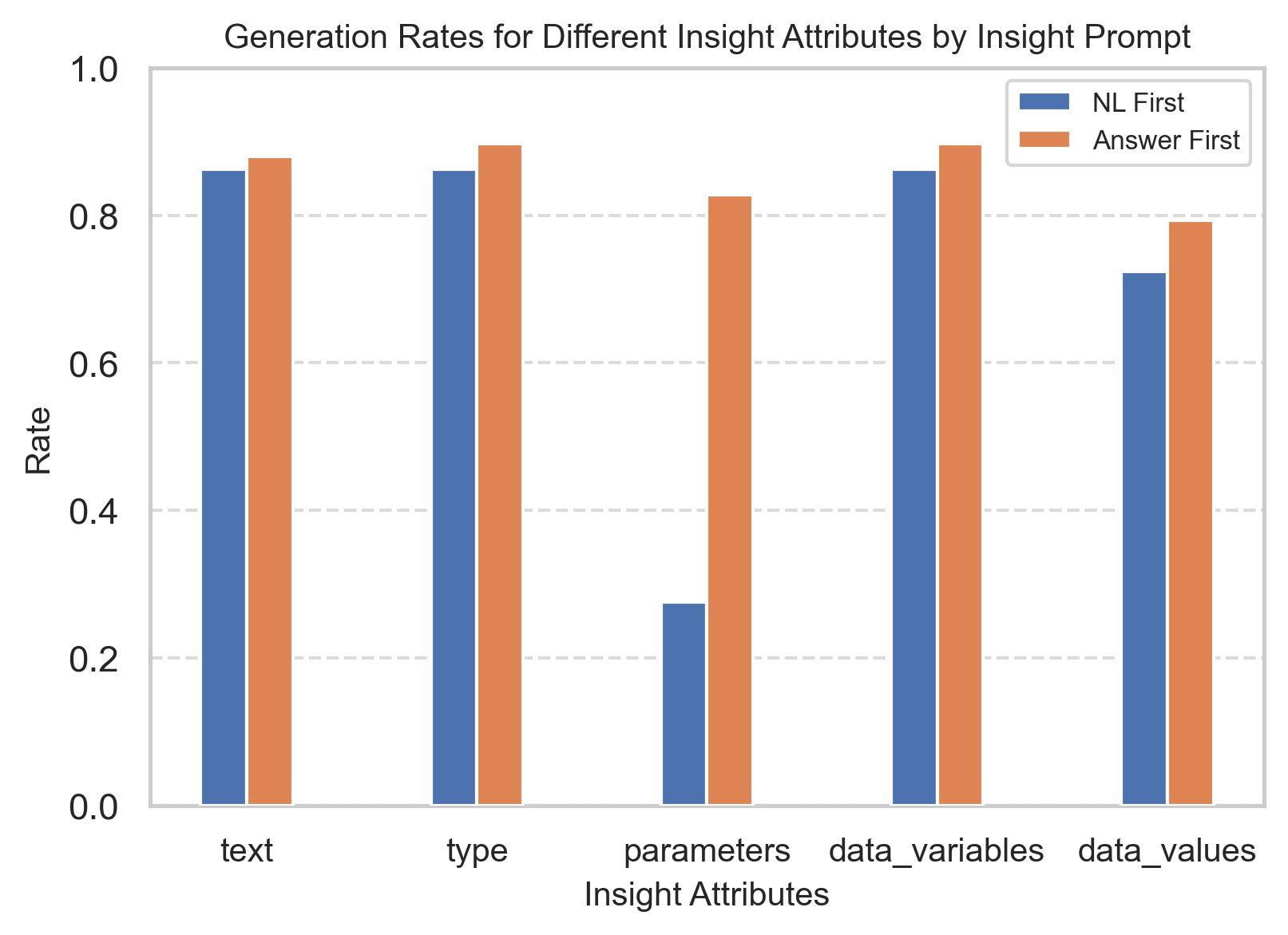}
    \vspace{-0.3cm}
    \caption{The insight attributes generation success rates (e.g., text, type, parameters) based on different prompts, with Answer first or Natural Language (NL) first.}
    \label{fig:insight_rate}
\end{figure}

\subsection{Time Cost}

To evaluate the efficiency of the models in terms of processing time, we measured the time costs for both code generation and insight generation tasks. From the point of view of time, GPT-3.5-turbo is significantly less than 4, which is 1/4 of the time of GPT-4-turbo. 

\begin{figure}[th]
    \centering
    \includegraphics[width=\linewidth]{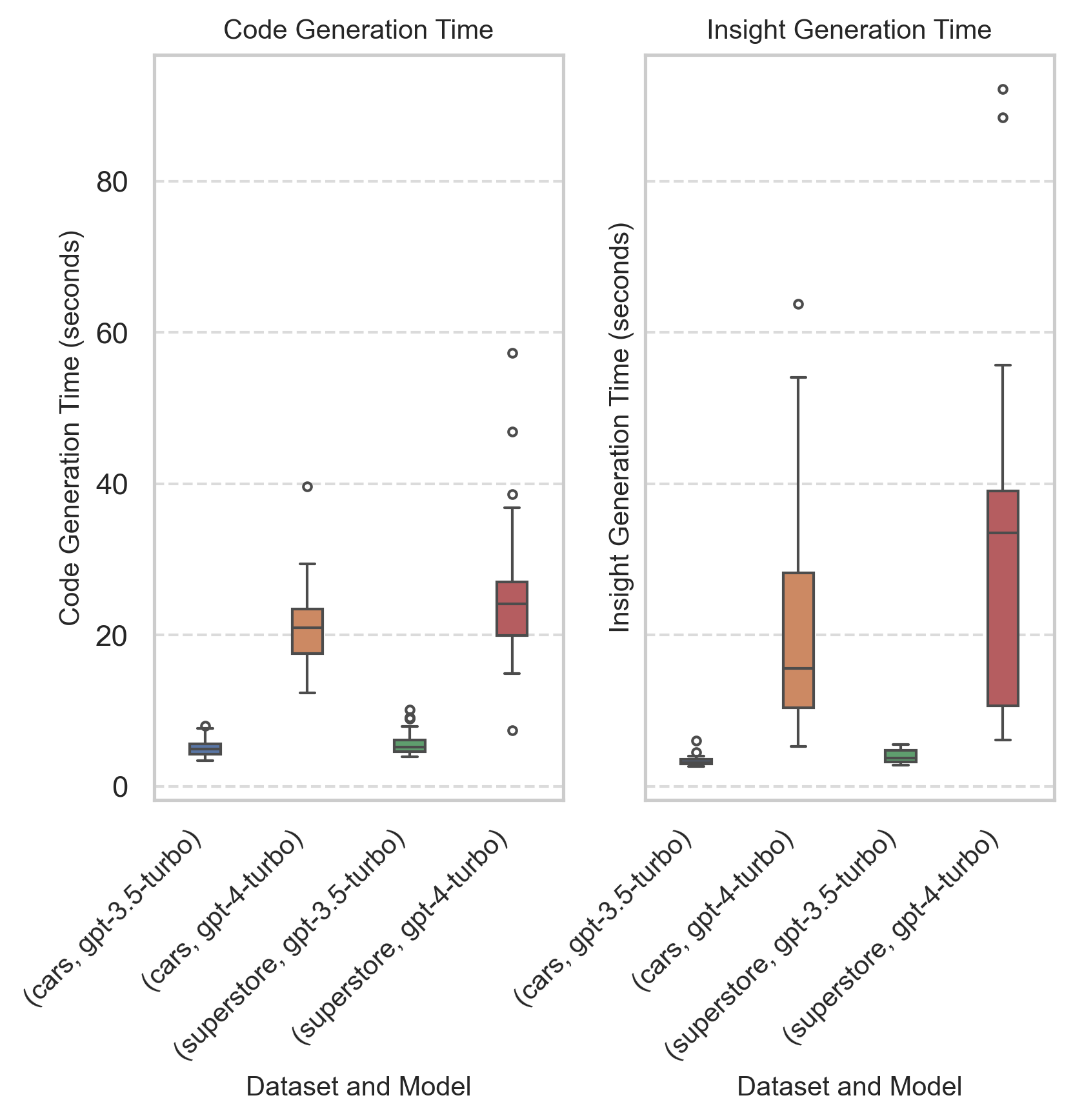}
    \vspace{-0.5cm}
    \caption{Time cost analysis for code and insight generation across different datasets (cars, superstore) and models (GPT-3.5-turbo, GPT-4-turbo), illustrating the variations in generation time based on model and dataset combination.}
    \label{fig:time_model}
    \vspace{-10pt}
\end{figure}

\textbf{Code generation time:} GPT-3.5-turbo demonstrates consistent performance with average times of 5.05 seconds for the cars dataset and 5.58 seconds for the superstore dataset, and relatively low variances of 1.32 and 2.34, respectively. In contrast, GPT-4-turbo has significantly higher average code generation times of 20.79 seconds for the cars dataset and 24.77 seconds for the superstore dataset, with much larger variances of 22.65 and 61.71, respectively, indicating greater variability and potentially less optimization.

\textbf{Insight generation time:} GPT-3.5-turbo again shows stable performance with average times of 3.36 seconds for the cars dataset and 3.94 seconds for the superstore dataset, and low variances of 0.44 and 0.70, respectively. On the other hand, GPT-4-turbo has much higher average insight generation times of 21.39 seconds for the cars dataset and 32.43 seconds for the superstore dataset, with significantly larger variances of 216.26 and 558.72, respectively, suggesting substantial variability and a need for further optimization.

Overall, GPT-3.5-turbo demonstrates more efficient and consistent performance in both code and insight generation tasks across different datasets.
GPT-4-turbo, while potentially more powerful, shows higher time costs but greater variability, suggesting it may be more sensitive to the complexity of tasks and datasets.